\documentclass[aps,prl,twocolumn,groupedaddress]{revtex4-2}
\usepackage{graphicx}
\usepackage{amsmath}
\usepackage{ifthen}
\usepackage{tabularx}
\usepackage{hhline}
\usepackage{amsfonts,amssymb,amsthm}
\usepackage{indentfirst}
\usepackage{parskip}

\begin{document}
\title{Nonlinear Optical Probing of Ferroic-Octupolar Order Parameter \\ in Collinear Altermagnet}

\author{P.~A.~Usachev}
\affiliation{Ioffe Institute, St. Petersburg, 194021, Russia}
\author{R.~V.~Pisarev}
\affiliation{Ioffe Institute, St. Petersburg, 194021, Russia}
\author{V.~V.~Pavlov}
\affiliation{Ioffe Institute, St. Petersburg, 194021, Russia}

\date{\today}

\begin{abstract}
Altermagnetism as a new concept in condensed matter physics is currently being thoroughly investigated. 
Despite the absence of macroscopic magnetization, altermagnets host a hidden spin order whose direct detection remains a challenge.
Here we report on the observation of electric and magnetic dipole forbidden optical second-harmonic generation (SHG) in the altermagnet CoF$_2$ with a centrosymmetric lattice and spin order. We demonstrate that below the N\'eel temperature $T_N = 38$~K the SHG signal is sensitive to the ferrotype magnetic octupole $\mathbf{\mathcal{O}}^M$ which is the order parameter in the antiferromagnetic phase.
By combining polarization-resolved SHG experimental data and a phenomenological symmetry analysis, we show that the altermagnetic spin structure of CoF$_2$ enables a ferroic-octupole-induced electric-quadrupolar nonlinear polarization $\mathbf{P}^{2\omega} = \mathrm i\varepsilon_{0}{ }^c\mathbf{\chi}^{(3)}(\mathbf{\mathcal{O}}^M) :\mathbf{E}^{\omega} \nabla \mathbf{E}^{\omega}$. The temperature dependence of SHG reveals a phase transition at $T_N$ confirming the spin origin of the observed signal. The SHG response 
is resonantly enhanced by a coherent three-photon process caused by the electronic $d$-$d$ transitions of the Co$^{2+}$ ion.
Model calculations of SHG polarization rotational anisotropies and temperature dependencies give a reasonable agreement with experimental data, proving the disclosed nonlinear contribution. Our results establish SHG as a novel sensitive tool of ferroic-octupolar spin ordering and highlight the potential of CoF$_2$ and other altermagnets for further nonlinear optical investigations and applications.
\end{abstract}
\maketitle

\emph{Introduction.} Altermagnets, recently identified as a distinct class of materials, exhibit a fundamental combination of properties characteristic of both antiferromagnets (AFM's) and ferromagnets \cite{Smejkal2022_0,Smejkal2022_1}. 
Although these materials are AFM's with fully compensated spin arrangement and no net magnetization, they host a spin-split electronic band structure that emerges even without atomic spin-orbit coupling \cite{Hayami}.
This momentum-dependent spin splitting, first studied in Refs.~\cite{Hayami,Yuan,Šmejkal}, gives rise to ferromagnet-like phenomena such as spin-polarized transport, spin-transfer torque, anomalous Hall effect, and unique spin–orbit and topological features \cite{Qu,Ma,Sattigeri}.
Such effects originate from the specific electronic structure of altermagnets, where strong spin-dependent states are linked to the wave vector and the order parameter \cite{Smejkal2022_0,Smejkal2022_1,Bhowal,Bai,Tamang,Jungwirth,Mostovoy}.
Examples of altermagnets include wide range of oxides, fluorides, chalcogenides, or intermetallic compounds \cite{Smejkal2022_1,Bhowal,McClarty,Adamantopoulos,Radaelli,Wei}.
Critically, this altermagnetic order parameter remains hidden from standard magnetometry and most linear optical techniques because of the absence of a net magnetization.
Developing a direct experimental approach to access this hidden spin texture is therefore a key objective in the field, making the search for optical phenomena capable of detecting this type of spin ordering a highly relevant task.

The optical second-harmonic generation (SHG) is a sensitive tool of symmetry breaking in both the charge \cite{Bloembergen,Shen,Boyd} and magnetic subsystems \cite{review,Fiebig1,Xiao}. Analysis of their symmetry properties enables separation of contributions from the crystal lattice and spin ordering, making the SHG a promising probe of hidden orders and phase transitions in different classes of antiferromagnets and multiferroics \cite{Fiebig2000,Zhao,Wang,Jin,Ahn}.
In centrosymmetric crystals, which include most altermagnets, the linear electric-dipolar contribution to SHG is strictly forbidden. However, higher-multipolar responses may become allowed.
As shown in \cite{Bhowal}, the ferrotype magnetic octupole $\mathbf{\mathcal{O}}^M$ serves as the order parameter below $T_N$ in the rutile-type cobalt difluoride CoF$_2$ which is the d-wave altermagnet \cite{Guo,Bhowal,Radaelli,CoF2_ref,Cheong}. CoF$_2$ is currently an intensively studied model AFM crystal \cite{Disa,Metzger,Dubrovin,Formisano}.

In this Letter we demonstrate electric and magnetic dipole forbidden SHG in the centrosymmetric altermagnet CoF$_2$, revealing a purely spin response originating from the ferrotype magnetic octupole $\mathbf{\mathcal{O}}^M$. 
Phenomenologically the spin-induced contribution to SHG is described by components of a $c$-type nonlinear susceptibility tensor that are odd under the time-reversal symmetry operation $\mathcal{T}$ \cite{Birss}. 
It emerges exclusively below the Néel temperature $T_N$, adding to the crystallographic background described by a $\mathcal{T}$-invariant $i$-type tensor. The appearance of a $\mathcal{T}$-noninvariant $c$-type contribution in SHG is a direct consequence of the ferrotype ordering of magnetic octupoles in CoF$_2$ \cite{Bhowal}.
The polarization anisotropy and the temperature evolution across $T_N = 38$~K are fully consistent with an octupolar origin of the SHG response. 
Our finding establishes the SHG as an effective method for the
direct probing of the ferroic-octupolar order in collinear
altermagnets, opening new avenues for their investigations and applications in AFM spintronics \cite{Baltz,Nemec}.

\emph{Results.} To investigate the nonlinear response of the altermagnetic CoF$_2$ crystal, we employed a femtosecond second-harmonic generation (SHG) technique \cite{Pavlov}.
Measurements were carried out in two polarization geometries ($E^{\omega} \parallel E^{2\omega}$ and $E^{\omega} \perp E^{2\omega}$) over the temperature range 1.6–50~K.
The $xz$-cut CoF$_2$ single crystal platelet was tilted by $30^\circ$ about the in-plane $z$-axis, which remained perpendicular to the incident fundamental beam ($\hbar\omega = 1.05$~eV). This oblique alignment is necessary to activate nonlinear susceptibility tensor components that are inaccessible under normal incidence. A complete description of the SHG experimental setup is provided in the Supplementary Information (SI 1).

\begin{figure}[ht]
    \centering
\includegraphics[width=0.46\textwidth]{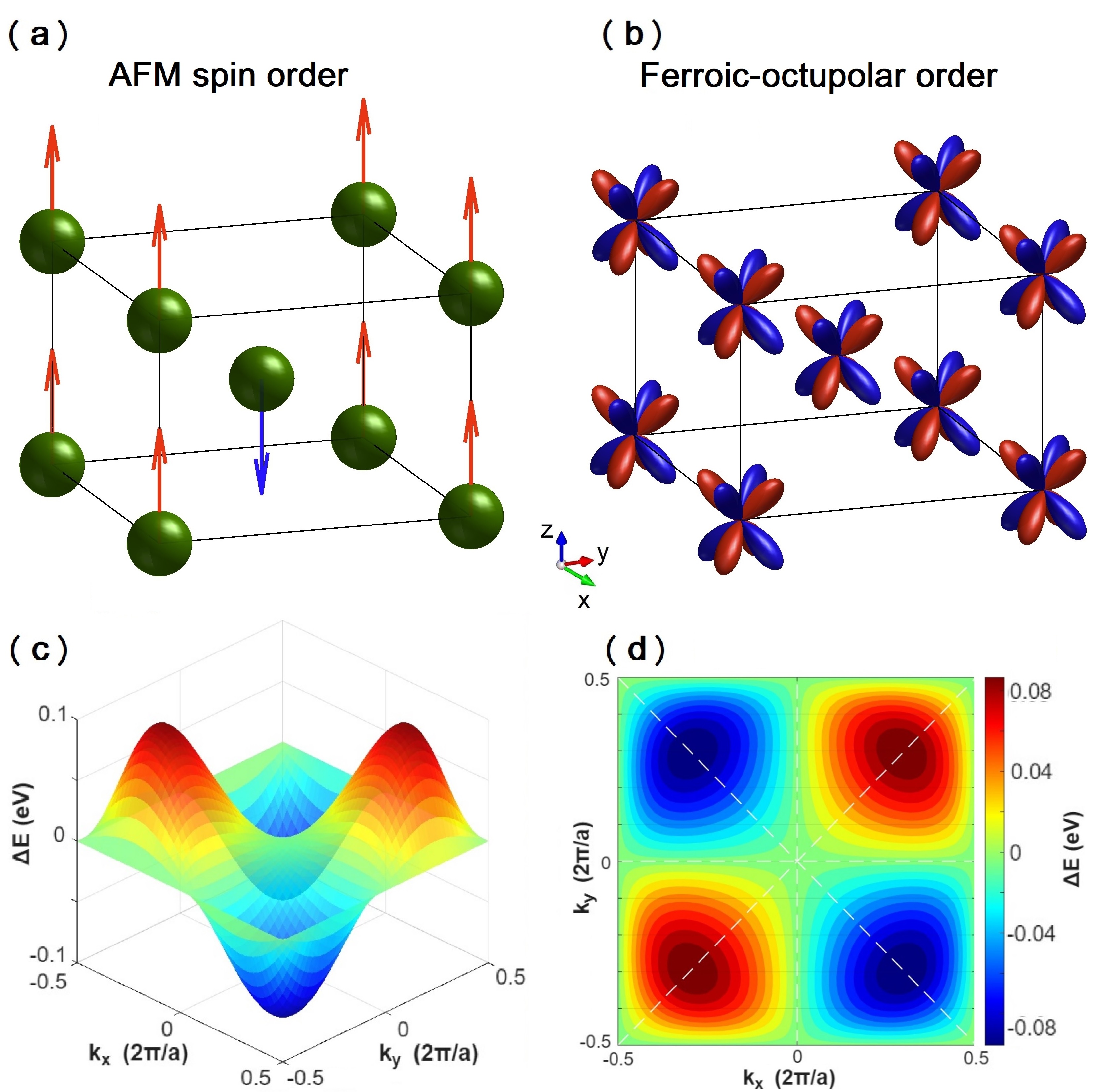}
    \caption{(a) The AFM spin order of Co$^{2+}$ ions and (b) the ferrotype ordering of magnetic octupoles in CoF$_2$. The angular distributions of magnetization density are marked by the red and blue colors representing regions of up and down spin polarizations, respectively~\cite{Bhowal}. (c) The 3D graph and (d) the flat mapping of d-wave spin splitting in CoF$_2$ (see SI 2 for details).}
    \label{fig1}
\end{figure}

Cobalt fluoride CoF$_2$ crystallizes in the rutile type structure with the space group $P4_2/mnm$ (point group $4/mmm$).
It develops a collinear AFM order below the Néel temperature $T_N = 38$~K, with spins directed along the $z$-axis \cite{Lines,Cowley}.
The corresponding AFM configuration of CoF$_2$ is shown in Fig.~\ref{fig1}a.
The magnetic symmetry (point group $4'/mmm'$) identifies CoF$_2$ as the $d$-wave altermagnet \cite{Guo,Bhowal,Radaelli,CoF2_ref,Cheong} with the ferrotype magnetic octupole $\mathbf{\mathcal{O}}^M$ acting as the order parameter \cite{Bhowal}. Figure~~\ref{fig1}b shows the angular distribution of the
magnetization density related to this type ordering of magnetic octupoles.
A direct consequence of this ferrotype octupolar order is the strong exchange spin splitting, that depends anisotropically on the wave-vector direction $\mathbf{k}$ (see Fig.~\ref{fig1}c,d), a most important feature of altermagnetic materials \cite{Smejkal2022_0,Smejkal2022_1,Bhowal,Radaelli,Cheong,Bai,Tamang,Jungwirth}. Calculation of spin splitting in CoF$_2$ is detailed in SI 2.

\begin{figure}[htbp]
\includegraphics[width=0.5\textwidth,{angle=0}]{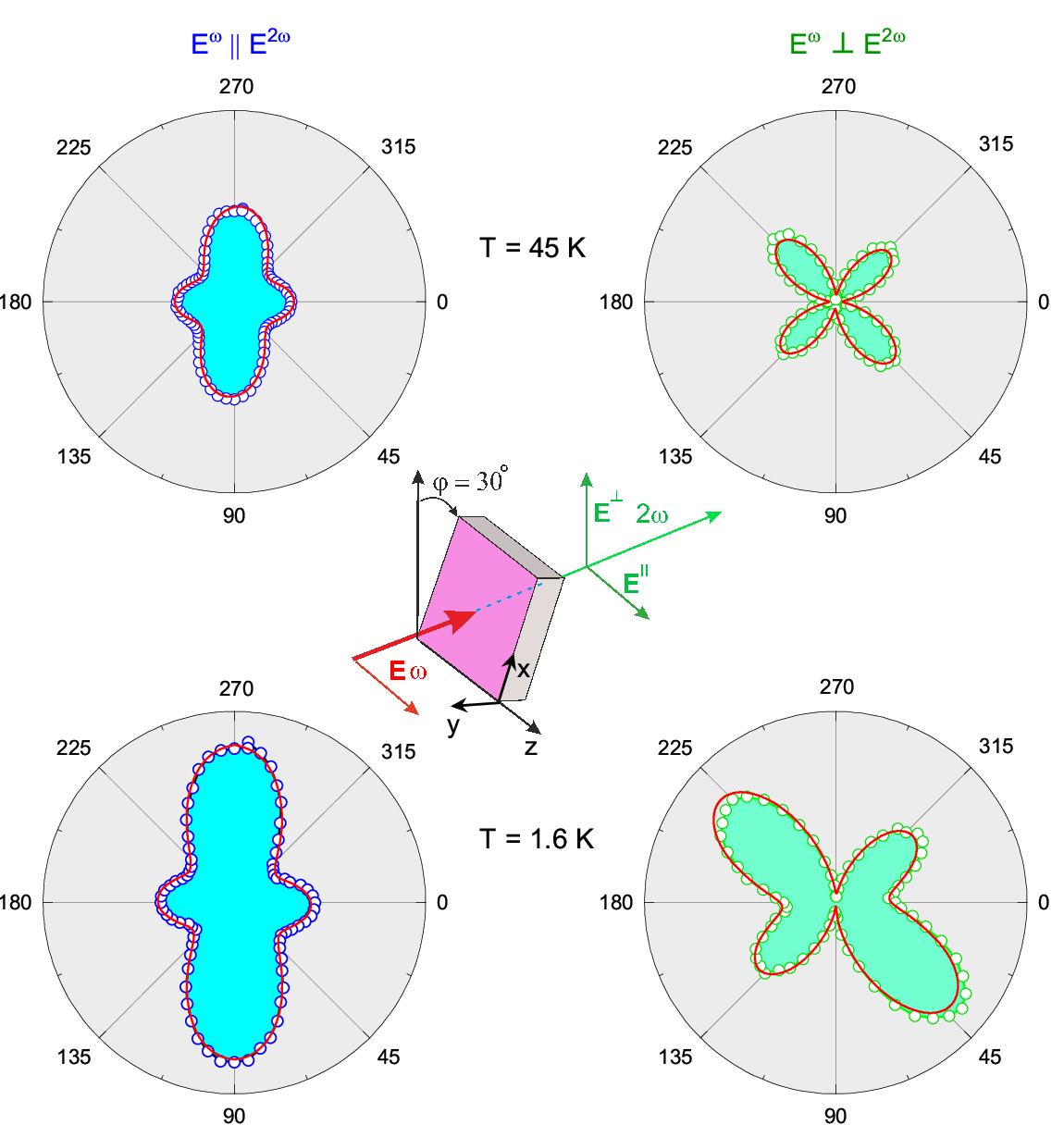}
\caption{Polarization-dependent SHG signals registered in the two geometries $E^{\omega} \parallel E^{2\omega}$ and $E^{\omega} \perp E^{2\omega}$ and at temperatures of $1.6$~K and $45$~K. The experimental SHG intensities are indicated by points, modelled SHG signals are shown by lines. Inset shows the parallel $E^{\omega} \parallel E^{2\omega}$ and crossed $E^{\omega} \perp E^{2\omega}$ light-polarization geometries.}
\label{fig2}
\end{figure}

To compare nonlinear responses in the paramagnetic and AFM phases, polarization-rotation anisotropies of SHG intensity were measured above and below $T_N$, for
$T = 45$~K and $1.6$~K, respectively, as shown in Fig.~\ref{fig2}.
In the paramagnetic phase the SHG signal is purely crystallographic, and it is governed by the $\mathcal{T}$-invariant $i$-type susceptibility tensor.
These results manifest as twofold and fourfold azimuthal patterns for the parallel ($E^{\omega} \parallel E^{2\omega}$) and crossed ($E^{\omega} \perp E^{2\omega}$) polarization geometries, respectively.
Crucially, below $T_N$, the SHG response undergoes a distinct transformation.
In the crossed geometry the emerging signal changes from the fourfold-type of polarization-rotation anisotropy ($T = 45$~K, above $T_N$) to the twofold-type ($T = 1.6$~K, below $T_N$), while in the parallel geometry the signal remains twofold-type at both temperatures. However, its intensity rises significantly at the low temperature. These qualitative changes in the SHG intensity point to an additional contribution governed by the spin order below $T_N$.

\begin{figure}[htbp]
\includegraphics[width=0.47\textwidth,{angle=0}]{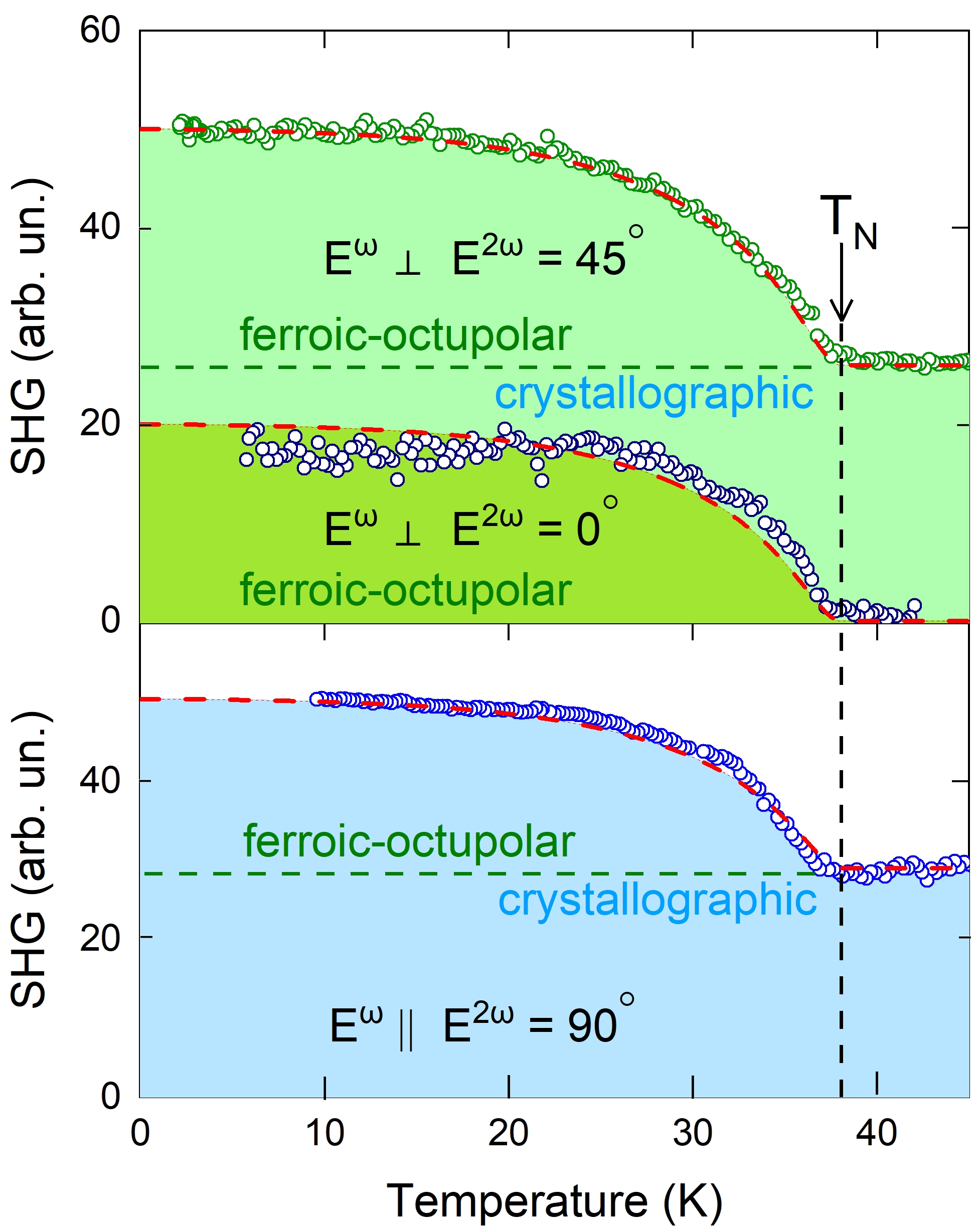}
\caption{Temperature dependencies of SHG intensity for two geometries
 $E^{\omega} \perp E^{2\omega}$ and $E^{\omega} \parallel E^{2\omega}$ (dots). The red dash lines show temperature dependencies of the squared order parameter of CoF$_2$ obtained from the neutron spectroscopy data \cite{Chatterji}.}
\label{fig3}
\end{figure}

Figure~\ref{fig3} shows the temperature evolution of SHG intensities in two polarization geometries for different angles ($E^{\omega} \perp E^{2\omega} = 0^\circ$ and $45^\circ$, $E^{\omega} \parallel E^{2\omega} = 90^\circ$), the angle is counted in respect to $E^{\omega}$ for both geometries. These experimental SHG data provide a quantitative measure of the spin-induced contribution.
A purely magnetic response is observed when the fundamental polarization is set to $0^\circ$ for the crossed geometry $E^{\omega} \perp E^{2\omega} = 0^\circ$: the signal is absent above $T_N$ and emerges only below the Néel temperature, increasing monotonously upon further cooling -- faithfully mirroring the temperature dependence of the AFM order parameter.
In contrast, the signal in crossed geometry $E^{\omega} \perp E^{2\omega} = 45^\circ$ contains both spin-induced and crystallographic contributions, as evidenced by its finite value above $T_N$. The same temperature behaviour is observed for the parallel geometry $E^{\omega} \parallel E^{2\omega} = 90^\circ$.

Therefore, the appearance of a new SHG component below $T_N$ and its distinct temperature dependence provide clear evidence that this component is tied to spontaneous AFM spin ordering.
As we show below, important that a specific phase relation of the crystallographic and spin-induced contributions renders the observed SHG signal insensitive to the distribution of AFM domains.
To establish the nature of this signal and the corresponding nonlinear susceptibility tensor, we now turn to a symmetry analysis of the observed SHG phenomena.

\emph{Discussion.}
A reasonable interpretation of the observed SHG signal, especially the extra contribution emerging below $T_N$, requires a systematic assessment of all symmetry-permitted nonlinear-optical mechanisms.
In a centrosymmetric crystal like CoF$_2$ (point group $4/mmm$) the electric-dipolar mechanism is strictly forbidden \cite{Bloembergen,Shen,Boyd}.
Within the magnetic-dipolar and electric-quadrupolar approximations, the crystallographic nonlinear optical polarization $\mathbf{P}^{2\omega}$ takes the form: \cite{Adler,Hoshi}:
\begin{equation} \label{eq1}
 \mathbf{P}^{2\omega} = \mathrm i\varepsilon_{0}{  }^i\mathbf{\chi}^{(2)} :\mathbf{E}^{\omega}\mathbf{H}^{\omega} + \mathrm \varepsilon_{0}{  }^i\mathbf{\chi}^{(3)} :\mathbf{E}^{\omega} \mathbf{Q}^{\omega},
\end{equation}
where, $\mathbf{E}^{\omega}$ and $\mathbf{H}^{\omega}$ are the electric and magnetic fields of the electromagnetic wave at the fundamental frequency $\omega$, respectively, $\mathbf{Q}^{\omega}$ is the electric quadruple at the fundamental frequency $\omega$, $\epsilon_0$ is the vacuum permittivity.
The first term in Eq.~(\ref{eq1}) is a magnetic-dipolar contribution. It is described by an axial third-rank tensor $^i\mathbf{\chi}^{(2)}$, which is formally allowed by symmetry. However, its specific tensor components ($yxz =-xyz$, $yzx = -xzy$, $zyx = -zxy$) alternate in sign, yielding a vanishing net contribution to SHG in CoF$_2$.

The second term in Eq.~(\ref{eq1}) is the crystallographic contribution to SHG of the electric-quadrupolar type.
This term reflects an anisotropic charge distribution and is therefore described by an $i$-type tensor $^i\mathbf{\chi}^{(3)}$.
It accounts for the background SHG signal observed, for instance, in the crossed geometry $E^{\omega} \perp E^{2\omega}= 45^\circ$ and in the parallel geometry $E^{\omega} \parallel E^{2\omega} = 90^\circ$, as shown in Fig.~\ref{fig3}.
Since its components are $\mathcal{T}$-invariant, or in other words insensitive to the spin order, they cannot explain the change in the SHG signal below $T_N$.
The onset of AFM order below $T_N$ breaks $\mathcal{T}$-symmetry, thereby enabling an additional $c$-type nonlinear susceptibility associated with the spin ordering \cite{Birss}. First, the SHG contribution of magnetic-dipole type $\mathrm i\varepsilon_{0}{ }^c\mathbf{\chi}^{(2)} :\mathbf{E}^{\omega}\mathbf{H}^{\omega}$ must be taken into account. The axial $\mathcal{T}$-odd tensor $^c\mathbf{\chi}^{(2)}$ has nonzero components ($zxy = zyx, xzy = yzx, xyz = yxz$). However, SHG signal below $T_N$ for the parallel geometry $E^{\omega} \parallel E^{2\omega} = 90^\circ$ cannot be described, since the tensor $^c\mathbf{\chi}^{(2)}$ for this geometry gives zero SHG output.  
Second, the SHG contribution of electric quadrupole type must be taken into account.
This contribution, which changes sign under time reversal, must be included in the description of the overall SHG signal.
Rewriting the electric-quadrupolar term in Eq.~(\ref{eq1}) to explicitly contain $\nabla\mathbf{E}^{\omega}$ \cite{Hoshi} yields the total nonlinear polarization $\mathbf{P}^{2\omega}$ as a superposition of crystallographic ($i$-type) and spin-driven ($c$-type) terms:
\begin{equation} \label{eq2}
 \mathbf{P
 }^{2\omega} = \mathrm \varepsilon_{0}{  }^i\mathbf{\chi}^{(3)} :\mathbf{E}^{\omega} \nabla \mathbf{E}^{\omega} + \mathrm i\varepsilon_{0}{  }^c\mathbf{\chi}^{(3)}:\mathbf{E}^{\omega} \nabla \mathbf{E}^{\omega}.
  \end{equation}

The specific structure of tensors $^i\mathbf{\chi}^{(3)}$ and $^c\mathbf{\chi}^{(3)}$ follows from the crystal and magnetic symmetries of CoF$_2$ below $T_N$.
In the AFM phase, the magnetic point group $4'/mmm'$ breaks $\mathcal{T}$-symmetry but preserves the combined $\mathcal{\rho T}$ operation ($\mathcal{\rho}=90^\circ$ rotation)  \cite{Radaelli,Kimel,Qu}. 
The nonzero independent components of $^i\mathbf{\chi}^{(3)}$ and $^c\mathbf{\chi}^{(3)}$, as derived from the magnetic point group $4'/mmm'$ in Ref.~\cite{Birss}, are presented in Table~\ref{tab:table01}.
The sign alternation seen in the $c$-type tensor (for instance, $^c\chi_{xxxx} = -^c\chi_{yyyy}$) directly reflect the $4'/mmm'$ symmetry and provide a clear marker of the spin-induced SHG. Further symmetry considerations of SHG are elaborated in the SI 3.

The $^c\mathbf{\chi}^{(3)}$ tensor in the spin-driven term of Eq.~(\ref{eq2}), governing the SHG signal below T$_N$, must be proportional to the AFM order parameter.
A proper candidate for such a parameter in CoF$_2$ is the ferrotype magnetic octupole $\mathbf{\mathcal{O}}^M$, the lowest-order magnetic multipole allowed by the $\mathcal{\rho T}$ symmetry of this $d$-wave altermagnet \cite{Bhowal}.
Explicitly, $\mathcal{O}^M{ijk} = \int r_i r_j m_k (\mathbf{r}) d^3r$, where $\mathbf{m}(\mathbf{r})$ is the magnetization density and $\mathbf{r}$ the position vector (see \emph{Appendix A}, describing the Landau theory and order parameters in altermagnets). This inversion-symmetric third-rank tensor constitutes the first nonvanishing net magnetic multipole in AFM's, like MnF$_2$ and CoF$_2$ \cite{Bhowal}.
In these type AFM's the opposite-spin ion environments give rise to the ferrotype ordering of magnetic octupoles despite the absence of net magnetization.
The ferrotype magnetic octupole $\mathbf{\mathcal{O}}^M$, being the AFM order parameter, underpins a range of phenomena such as the piezomagnetism, second-order magnetoelectric effect \cite{Spaldin2022}, nonrelativistic spin splitting \cite{Bhowal}, and, as demonstrated in this work, the ferroic octupolar SHG.

Expressing the electric-quadrupolar terms via $\nabla\mathbf{E}^\omega$ in Eq.~(\ref{eq2}) allows us to explicitly incorporate the dependence on the light wave vector $\mathbf{k}^0$, since $\nabla\mathbf{E}^\omega \propto i \mathbf{E}^\omega \mathbf{k}^0$ (see SI 3).
This formulation naturally accounts for spatial dispersion effects \cite{Agranovich} and is particularly relevant for altermagnets such as CoF$_2$, where the electronic bands exhibit a strong $\mathbf{k}$-dependent spin splitting (Fig.~\ref{fig1}c) \cite{Guo,Bhowal,CoF2_ref,Cheong}. 
The discussion of the spin splitting and spatial dispersion effect in CoF$_2$ is also provided in SI 2 and 3.

\begin{table}[h]
\caption{\label{tab:table01} Nonvanishing components of nonlinear susceptibilities $^i\mathbf{\chi}^{(3)}$ and $^c\mathbf{\chi}^{(3)}$ for the crystallographic point group $4/mmm$ and magnetic point group $4'/mmm'$ \cite{Birss}. These tensors are symmetric with respect to the permutation of their second and fourth indices.}

\begin{tabular}{ccc}\hhline{=~~}
$^i\chi_{ijkl}$ ($4/mmm$, $4'/mmm'$), 8 independent components: \\ 
$xxxx= yyyy$, $yxxy= xxyy$, $yxyx= xyxy$,\\
$yyzz= xxzz$, $yzyz= xzxz$, $zyyz= zxxz$, \\
$zyzy= zxzx$, $zzzz$ \\ \hline
$^c\chi_{ijkl}$ ($4'/mmm'$), 7 independent components: \\ 
$xxxx=-yyyy$, $yxxy=-xxyy$, $yxyx=-xyxy$, \\
$yyzz=-xxzz$, $yzyz= -xzxz$, \\
$zyyz= -zxxz$, $zyzy= -zxzx$\\ \hline
\end{tabular}
\end{table}

A model based on the phenomenological Eq.~(\ref{eq2}) and the tensor components of Table~\ref{tab:table01} (solid lines in Fig.~\ref{fig2}) quantitatively reproduces the measured polarization anisotropies for the parallel and crossed geometries. 
The fit requires the coexistence of $i$- and $c$-type contributions below $T_N$, while above $T_N$ the $i$-type crystallographic term alone is sufficient. 
The validity of this description is further supported by the strong suppression of the SHG signal at normal incidence (Fig.~3, SI 3). 

The observed temperature dependencies of SHG intensity below $T_N$ follow the temperature behavior of the squared AFM order parameter of CoF$_2$ obtained from the neutron spectroscopy data \cite{Chatterji} (see \emph{Appendix B}, describing relation of the order parameter $\mathbf{\mathcal{O}}^M$ with $\mathcal{T}$–odd SHG tensor $\text{ }^c\mathbf{\chi}^{(3)}$ and its temperature dependence). 
Important to say that the $\pi/2$ phase shift between $i$- and $c$-type contributions prevents interference, making the signal insensitive to AFM domain structure.
Fitting the SHG intensity as $I^{2\omega} \propto \big |\big(^i\mathbf{\chi}^{(3)}+ i\text{ }^c\mathbf{\chi}^{(3)}(\mathbf{\mathcal{O}}^M)\big)I^{\omega}\big |^2$ within this framework reproduces well the experimental data, providing a complete proof that the ferrotype magnetic octupole $\mathbf{\mathcal{O}}^M$ is the AFM order parameter underlying the spin-induced SHG in CoF$_2$. 
The tensor relation linking the ferroic octupolar components of $^c\mathbf{\mathcal{O}}^M$ to $^c\mathbf{\chi}^{(3)}$ is detailed in SI 4, the temperature and polarization dependencies of SHG are modeled in SI 5.

\begin{figure}[htbp]
\includegraphics[width=0.48\textwidth]{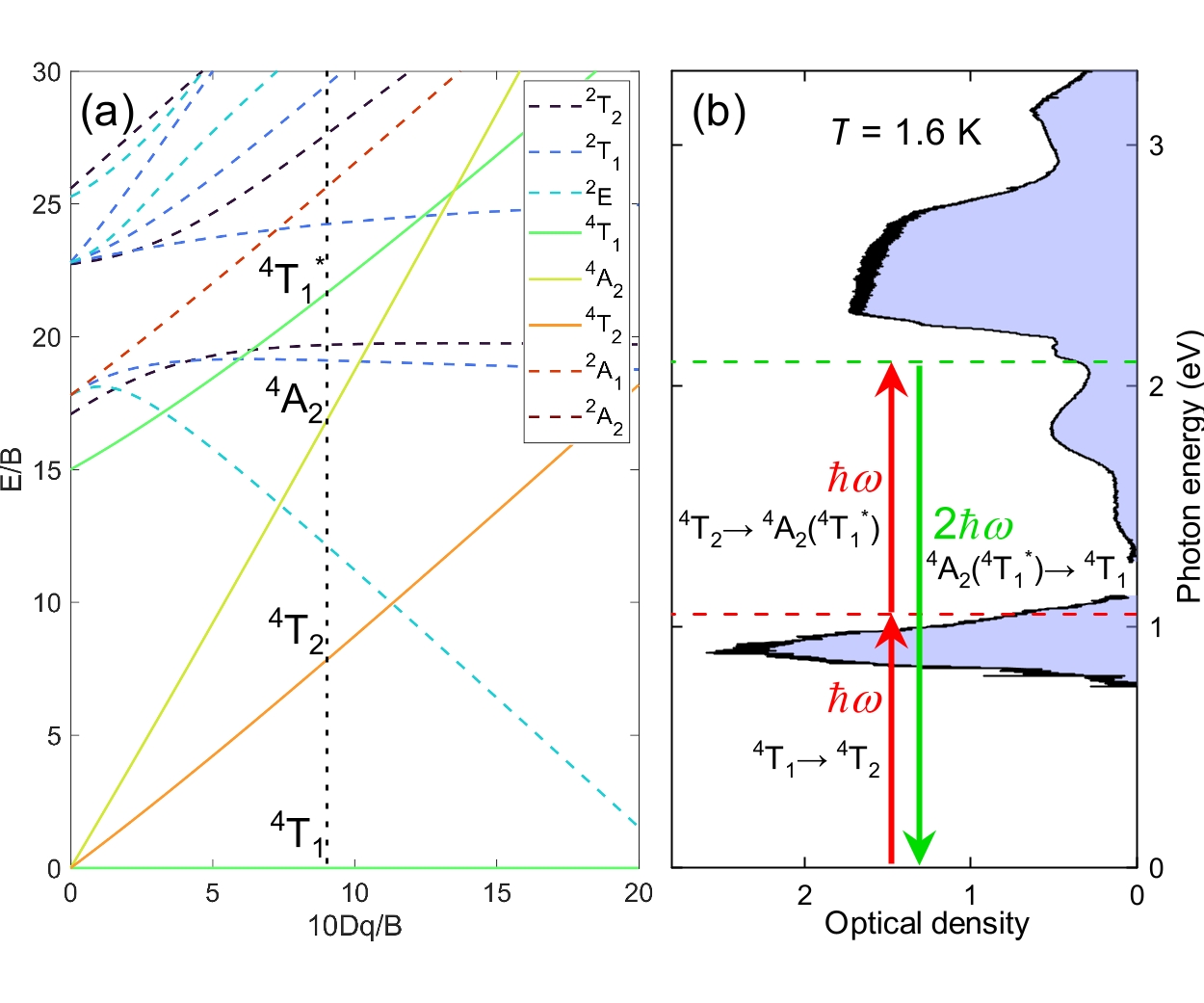}
\caption{(a) Tanabe-Sugano diagram for 3d$^7$ electron complex of the Co$^{2+}$ ion in the octahedral crystal field of F$^{-}$ ions, $10Dq$ is the crystal field splitting parameter, $B$ is the Racah parameter, $E$ is the photon energy. The vertical dot line is placed at the value of $10Dq/B = 9$ ($10Dq = 0.962$~eV, $B = 0.107$~eV)   \cite{Barreda}. Spin-allowed and spin-forbidden transitions are indicated by solid and dashed lines, respectively. (b) Optical
absorption spectrum of CoF$_2$ at $T = 1.6$~K measured for the unpolarized light. The vertical up arrows mark two photons with the fundamental energies $\hbar\omega$ and the down arrow marks a photon with the SHG energy $2\hbar\omega$.}
\label{fig4}
\end{figure}

The electric-quadrupolar mechanism responsible for both terms in Eq.~(\ref{eq2}) typically produces a weak nonlinear response (see \emph{Appendix C}, describing the SHG process and $\textbf{k}$-dependent spin splitting of electronic bands in CoF$_2$). 
To overcome this, we exploit resonant enhancement by selecting photon energies that align with spin-allowed $d$-$d$ transitions of the Co$^{2+}$ ion. Fig.~\ref{fig4}a  illustrates the Tanabe-Sugano diagram for 3d$^7$ electron complex of the Co$^{2+}$ ion in the octahedral crystal field of F$^{-}$ ions. A calculation of Tanabe-Sugano diagram \cite{Piccinni} is done for the ratio of Racah parameters $C/B=4.6$ \cite{Barreda}.
The absorption spectrum of CoF$_2$ (Fig.~\ref{fig4}b) features transitions from the $^4T_1$ ground state to $^4T_2$, ${ }^4A_2$, and $^{4}T^*_{1}$ excited states \cite{Ziel}.  
The SHG photon energy of $2\hbar\omega = 2.12$~eV coincides with a coherent three-photon process $^4T_1 \overset{\omega}{\rightarrow} {}^4T_2 \overset{\omega}{\rightarrow} {}^4A_2(^{4}T^*_{1}) \overset{2\omega}{\rightarrow} {}^4T_1$ (see Fig.~\ref{fig4}b).
All three constituent transitions are spin-allowed within the octahedral fluorine environment of Co$^{2+}$ ion.
This three-photon resonance effectively amplifies both the crystallographic $^i\mathbf{\chi}^{(3)}$ and the magnetic $^c\mathbf{\chi}^{(3)}$ nonlinear susceptibilities, allowing the otherwise subtle SHG signal to be measured with a high fidelity (see SI 6 for more details).

The classification of compensated antiferromagnets into altermagnets, magnetoelectrics, and conventional AFMs \cite{Kimel} provides a natural context for our findings. 
This study completes the SHG survey across these three classes and demonstrates that the AFM order parameter, the ferrotype magnetic octupole $\mathbf{\mathcal{O}}^M$ in CoF$_2$, directly governs the nonlinear optical response.
As summarized in Table~\ref{tab:table02}, the mechanism in CoF$_2$ is qualitatively distinct: unlike Cr$_2$O$_3$, where below $T_N$ SHG is electric-dipole-allowed \cite{Fiebig,Pavlov} via the magnetoelectric multipole $\mathcal{M}_{ij} = \int r_i m_j (\mathbf{r}) d^3r$ \cite{Bhowal,Spaldin}, or NiO, which exhibits $i$-type magnetic-dipolar SHG \cite{Fiebig2001}, below $T_N$ CoF$_2$ displays SHG driven by the octupolar order $\mathbf{\mathcal{O}}^M$ through a $\mathcal{T}$-odd $c$-type electric-quadrupolar susceptibility $^c\mathbf{\chi}^{(3)}(\mathbf{\mathcal{O}}^M)$.
Thus, the tabletop method of measuring SHG provides a direct, symmetry-selective probe of the AFM order parameter.

\begin{table}[h]
\caption{\label{tab:table02} Comparison of SHG mechanisms in AFM's.}
\begin{tabular}{c|c|c|c}\hhline{====}
Material    & Magnetic      & Symmetry in & SHG \\
            & class         & magn. phase & mechanism\\ \hline
CoF$_2$     & altermagnet   & centro- & electric quadrupole \\
$[$this work$]$  &               & symmetric               & $^c\mathbf{\chi}^{(3)}(\mathbf{\mathcal{O}}^M) :\mathbf{E}^{\omega} \nabla \mathbf{E}^{\omega}$ \\\hline
Cr$_2$O$_3$&magnetoelectr.& noncentro- & electric dipole \\
 \cite{Fiebig}&    AFM        &      symmetric          & $^c\mathbf{\chi}^{(2)}(\mathbf{\mathcal{M}}) :\mathbf{E}^{\omega} \mathbf{E}^{\omega}$ \\\hline
NiO         & conventional & centro- & magnetic dipole\\
 \cite{Fiebig2001}&    AFM        &      symmetric         & $^i\mathbf{\chi}^{(2)} :\mathbf{E}^{\omega} \mathbf{H}^{\omega}$ \\ \hline
\end{tabular}
\end{table}

\emph{Conclusions.} We demonstrate that second-harmonic generation in centrosymmetric CoF$_2$ is driven by the ferrotype magnetic octupole $\mathbf{\mathcal{O}}^M$ --
the AFM order parameter of $d$-wave altermagnetic state.
By combining polarization-resolved measurements with symmetry analysis, we identify a $c$-type tensor contribution, which emerges exclusively below $T_N$ as a result of $\mathcal{T}$-symmetry breaking and corresponds to the magnetic-octupole-induced electric-quadrupolar nonlinear polarization
$\mathbf{P}^{2\omega} = \mathrm i\varepsilon_{0}{ }^c\mathbf{\chi}^{(3)}(\mathbf{\mathcal{O}}^M) :\mathbf{E}^{\omega} \nabla \mathbf{E}^{\omega}$.
The temperature dependence of the SHG signal proves its purely magnetic origin and follows the temperature behavior of the squared AFM order parameter.
An important addition is that the coherent three-photon process $^4T_1 \overset{\omega}{\rightarrow} {}^4T_2 \overset{\omega}{\rightarrow} {}^4A_2(^{4}T^*_{1}) \overset{2\omega}{\rightarrow} {}^4T_1$ resonantly enhances the nonlinear susceptibility $^c\mathbf{\chi}^{(3)}(\mathbf{\mathcal{O}}^M)$.
Thus, we establish the SHG method as a tabletop sensitive optical technique for detecting the ferroic-octupolar order parameter.
This method is a versatile symmetry-selective tool for investigating hidden spin order, opening new avenues for nonlinear optical studies of altermagnetic materials.

\subsection*{ACKNOWLEDGMENTS}
This work was supported by the Russian Science Foundation (grant no. 24-12-00348). We thank R.~M.~Dubrovin for useful discussions.

\subsection*{Data availability}
The data supporting the conclusions in this Letter are publicly available~\cite{data}.

\textbf{\emph{End Matter}}

\emph{Appendix A: Landau theory and order parameters in altermagnets}

The thermodynamic Gibbs potential takes the form in the general Landau theory as~\cite{Landau,Schiff2025} 
\[
\Phi(\mathbf{N}) = a_2 (\mathbf{N}\cdot\mathbf{N}) + a_4 (\mathbf{N}\cdot\mathbf{N})^2 + \cdots,
\]
where $\mathbf{N}$ is a phenomenological order parameter, and $\Phi(\mathbf{N})$ is an analytic function of $\mathbf{N}$.
In the ordered phase, \(\mathbf{N}\neq\mathbf{0}\), and therefore it is the primary order parameter. 
In a two-sublattice AFM, the primary order parameter is the Néel vector $\mathbf{N} = \mathbf{m}_1-\mathbf{m}_2$, defined as the difference between the local magnetic moments on the sublattices.
Because idealized altermagnets lack spin–orbit coupling (SOC), the
symmetries that leave \(\mathbf{N}\) invariant belong to a spin-space group
\cite{Brinkman1966}. 
The approach based on the Néel vector \(\mathbf{N}\) can be used for phenomenology of altermagnets \cite{Mostovoy}. 
However, the Néel vector does not offer the same mathematical convenience as ferroic order parameters such as the magnetization in ferromagnets or the electric polarization in ferroelectrics~\cite{Bhowal}.

Ref.~\cite{Schiff2025} counts multipole order parameters by types \(n=0,\dots,6\). 
The magnetic multipole (MM) of type $n=0$ corresponds to the axial vector of magnetization $\mathbf{M} = \int \mathbf{m} (\mathbf{r}) d^3r$, which is zero in fully compensated AFM CoF$_2$. MM of type $n=1$ is related to the space-inversion-odd magnetoelectric quadrupole $\mathcal{M}_{ij} = \int r_i m_j(\mathbf{r}) d^3r$ which is an axial second-rank tensor. $\mathcal{M}_{ij}$ is zero in CoF$_2$ since it has centrosymmetric spin structure. MM of type $n=2$ corresponds to the inversion-symmetric magnetic octupole $\mathcal{O}^M_{ijk} = \int r_i r_j m_k(\mathbf{r}) d^3r$. This axial third-rank tensor describes spin order parameters in AFM's CoF$_2$ and MnF$_2$~\cite{Bhowal}. The magnetic octupole $\mathcal{O}^M$ is called as a secondary (or pseudoprimary) multipolar order parameter~\cite{Schiff2025}, giving rise to the $\textbf{k}$-dependent spin splitting of electronic bands.  

\emph{Appendix B: Relation of the order parameter $\mathcal{O}^M$ with $\mathcal{T}$--odd SHG tensor $^{c}\chi^{(3)}$ and its temperature dependence in CoF$_2$} 

As shown in Ref.~\cite{Schiff2025} certain MM order parameters
 are linearly related to the Néel
vector \(\mathbf{N}\). The Néel vector \(\mathbf{N}\) is described as a $\mathcal{T}$-odd axial vector with the representation $aeV$ in Jahn notation~\cite{Jahn1949}.
Considering SOC, MM's of mixed polar and magnetic character transform as $aeV \otimes [V^n]$.
The lowest nonvanishing MM in CoF$_2$ is magnetic octupole $\mathcal{O}^M$ of type $n=2$ with the representation $aeV[V^2]$. When SOC is included, the Néel vector \(\mathbf{N}\) couples linearly to $\mathcal{O}^M$~\cite{Schiff2025}. 

The coupling of SHG polarization \(\mathbf{P}^{2\omega}\) to \(\mathcal{O}^M\) is mediated through the magnetic-octupole-induced electric-quadrupolar interaction \(\mathbf{P}^{2\omega}= i \varepsilon_0 {}^{c}\chi^{(3)} (\mathcal{O}^M) \mathbf{E}^\omega \nabla\mathbf{E}^\omega \). 
The magnetic-octupole-induced SHG in CoF\(_2\) is described by a polar $\mathcal{T}$-odd tensor of fourth-rank 
\({}^{c}\chi^{(3)}\) transforming as \(aV^2[V^2]\).
The relationship between the nonlinear susceptibility ${}^{c}\chi^{(3)}$ and $\mathcal{O}^M$ by the axial $\mathcal{T}$-even tensor of sevens-rank $^i\chi_{ijklmno}$ ($eV^3[V^2][V^2]$) is discussed in SI 4. 
The nonzero components of \({}^{c}\chi^{(3)}\) (as well as other tensors discussed in this Letter) can be found in Birss tables~\cite{Birss}, on the Bilbao crystallographic server~\cite{Aroyo}, or using a specially developed Matlab program~\cite{data}.

In CoF\(_2\), the neutron scattering data
present the low-energy inelastic magnetic peak intensity~\cite{Chatterji}, that is
proportional to a sublattice magnetization \(\mathbf{m}(T)\) of CoF$_2$. 
Below the Néel temperature the AFM vector \(\mathbf{N}(T)\) and the magnetic octupole \(\mathcal{O}^M(T)\) have identic temperature dependencies as they are coupled linearly~\cite{Schiff2025}.
The SHG signal below \(T_N\) shows the squared
temperature dependence of inelastic magnetic peak intensity (see Fig.~3). SHG intensity is defined as $I^{2\omega}(T) \propto \mid\mathbf{P}^{2\omega}\mid^2 \propto [\mathcal{O}^M(T)]^2$, therefore the spin-induced SHG originates from the AFM ordering described by
the ferro-type magnetic octupole \(\mathcal{O}^M\).

\emph{Appendix C: SHG process and $\textbf{k}$-dependent spin splitting of electronic bands in CoF$_2$}

The SHG electric-quadrupolar contribution $^{c}\chi^{(3)} (\mathcal{O}^M) \mathbf{E}^\omega \nabla\mathbf{E}^\omega$ is
proportional to the light wave vector \(\mathbf{k}^0\), as \(\nabla\mathbf{E}^\omega \propto i\mathbf{k}^0\mathbf{E}^\omega\), see SI 3.
For a nonzero \(\mathbf{k}^0\), the electric-quadrupolar nonlinear interaction
involves electronic states near the \(\Gamma\) point in the Brillouin
zone. Model calculations show the
substantial spin splitting of \(\sim 0.1\)~eV at $k\approx 0.25$ (units $2\pi/a$, where $a$ is the lattice parameter), see SI 2. 
The spin splitting originates from an energy term of the
form $J_z\sigma_z \sin(k_x a) \sin(k_y a)$, where $J_z$ is the exchange integral and $\sigma_z$ denotes the spin operator~\cite{Sarkar}. For the wave vector \(k^0 \approx 0.008\) ($2\hslash\omega = 2.1~eV$) the spin splitting $\Delta E(\mathbf{k}) \propto k_x k_y$ \cite{Smejkal2022_0} is about $1.5$~meV, therefore the Kramers degeneracy is lifted in the Brillouin zone for SHG process probing CoF$_2$. 
Thought the spin splitting is small, the SHG process probes it through the nonlinear optical susceptibility \({}^{c}\chi^{(3)}\) determined by matrix elements of real electronic transitions, see SI 6. Hence, a resonance enhancement due to the three-photon resonance with \(d\)-\(d\) transitions at 2.1~eV (see Fig.~4) notably amplifies the SHG signal. Moreover, the presence of a nonzero \(\mathbf{k}^0\) breaks the inversion symmetry of the electronic wave functions at the microscopic level, allowing quadrupolar transitions that couple to the anisotropic \(\mathbf{k}\)-dependent spin splitting. Thus, though the absolute value of \(\mathbf{k}^0\) is small, the SHG signal is sensitive to the anisotropic spin splitting in the \(d\)-wave altermagnet CoF$_2$.

\end{document}


\maketitle

\section{Second-harmonic generation experimental setup}
\begin{figure}[ht]
    \centering
\includegraphics[width=0.8\textwidth]{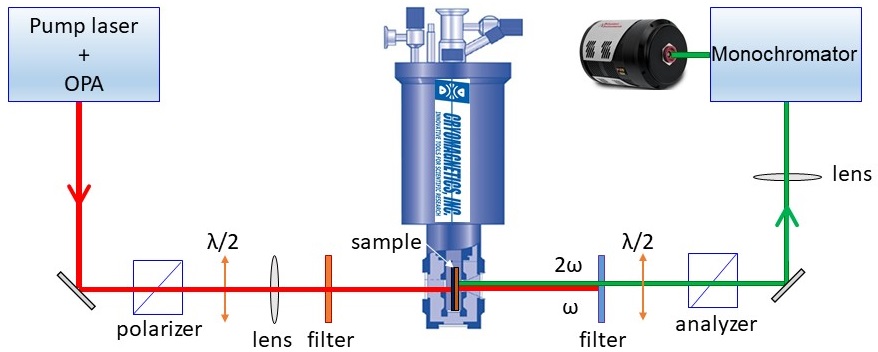}
\caption{ SHG experimental setup based on a femtosecond pump laser and an optical parametric amplifier (OPA).}
\label{setup}
\end{figure}
The optical second-harmonic generation (SHG) is measured by the experimental setup based on a femtosecond pump laser as shown in Fig.~\ref{setup}.
The laser system consists of
an optical parametric amplifier (OPA), pumped by a $1$~kHz pulsed laser,
with a tunable output across the spectral range of interest, and $\sim50$~fs pulse duration.
The OPA generates fundamental radiation (at frequency $\omega$) with photon energy $\hbar\omega = 1.05$~eV and pulse energies up to 6~$\mu$J. The fundamental beam is focused to a $\sim100~\mu$m diameter spot.
In experiments a high-precision ($<0.5^\circ$ alignment) $xz$-plane oriented CoF$_2$ sample ($z$ is the $c$-axis of the single crystal) was used. The sample was mounted in a helium-bath cryostat ($T = 1.6 - 50$~K). The polarizations of the fundamental and SHG (at frequency $2\omega$) beams were controlled using half-wave plates, and the SHG light was passed through a monochromator. The SHG signal was recorded using a cooled CCD photo-detector PIXIS 256E. Measurements were carried out for the parallel $E^{\omega} \parallel E^{2\omega}$ and crossed $E^{\omega} \perp E^{2\omega}$ geometries. For both geometries the azimuthal angle was counted for the $E^{\omega}$ polarization from the $z$-direction of the CoF$_2$ sample (see Inset in Fig.~2 in the main text). The SHG measurements were done at the photon energy  $2\hbar\omega = 2.1$~eV.

\section{Crystallographic and spin structures of CoF$_2$, ferroic-octupolar order and microscopic spin splitting}

\begin{figure}[ht]
 \centering
\includegraphics[width=.6\textwidth]{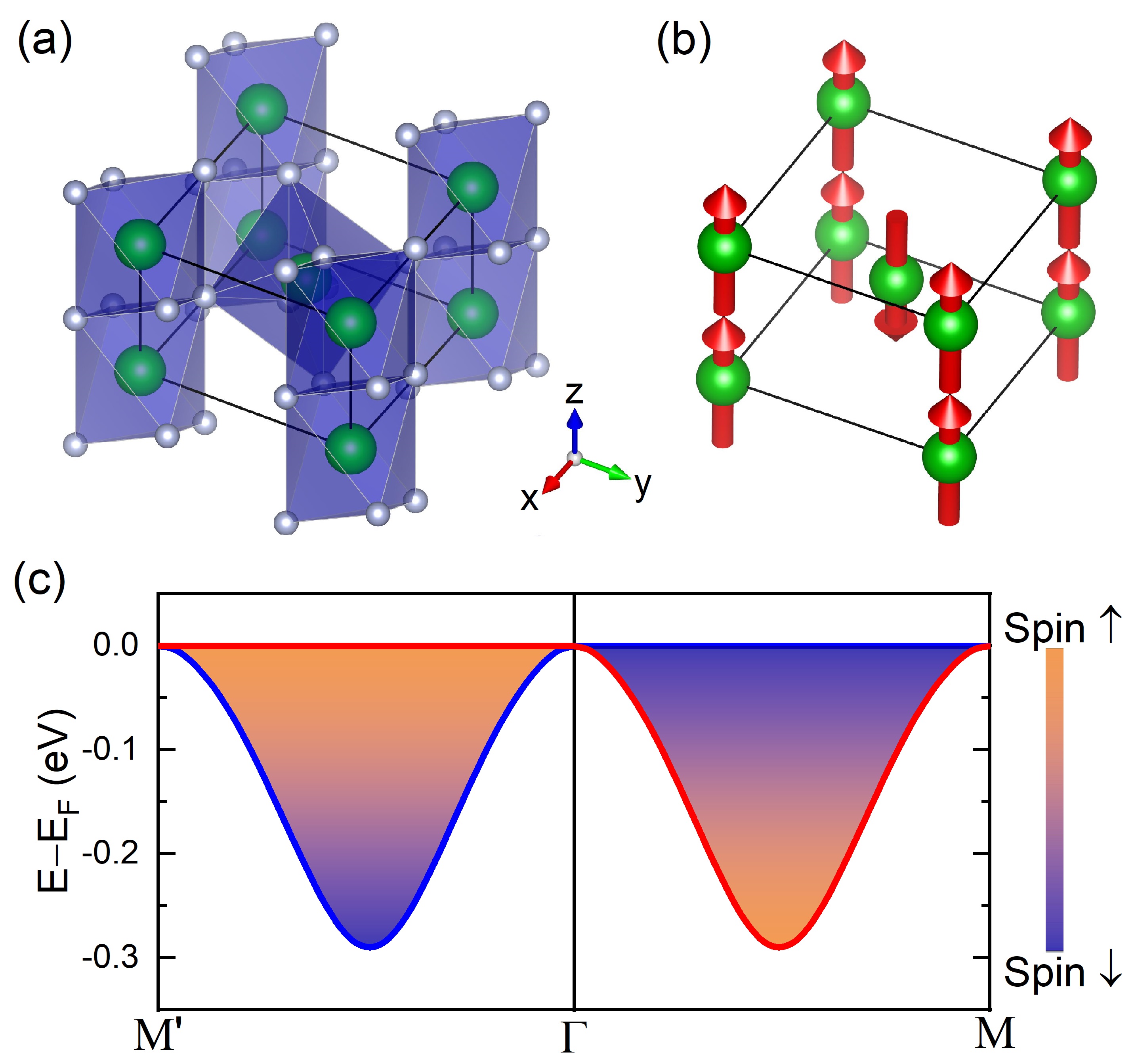}
\caption{(a) Crystallographic tetragonal structure of the CoF$_2$ with polyhedra denoting the octahedral environment of magnetic Co$^{2+}$ ions (green spheres) by F$^{-}$ ions (grey spheres), (b) spin structure of CoF$_2$ below $T_N$ with spins oriented along the tetragonal $z$-axis. Red arrows show spins of Co$^{2+}$ ions for a single AFM domain
of the unit cell, visualized with VESTA \cite{VESTA}. (c) The energy within $k$-space (Brillouin zone) and the spin splitting $\uparrow$ and $\downarrow$ of electronic bands ($k_x$ -- $\Gamma$-M, $k_y$ -- $\Gamma$-M$'$) in CoF$_2$ \cite{Guo,CoF2_ref} performed by the Vienna \emph{ab-initio} simulation package software (VASP).}
\label{fig2}
\end{figure}

Cobalt difluoride CoF$_2$ crystallizes in the tetragonal rutile structure (crystallographic space group $P4_2/mnm$) with lattice parameters $a = \SI{4.6951}{\angstrom}$ and $c = \SI{3.1796}{\angstrom}$ \cite{Stout}. The unit cell contains two Co$^{2+}$ ions at Wyckoff positions 2a: (0, 0, 0) and (1/2, 1/2, 1/2), as  illustrated in Fig.~1a. 
Above the Néel temperature $T_N = 38$~K, the symmetry is described solely by this crystallographic space group. Upon cooling below $T_N$, the system undergoes a transition to a collinear antiferromagnetic (AFM) state with spins aligned along the $z$-axis, see Fig.~1b.
The spin order lowers the symmetry, which is described by the magnetic (Shubnikov) space group $P4_2'/mnm'$. Here, the primes signify that certain spatial symmetry operations are combined with the time-reversal symmetry operation $\mathcal{T}$. This magnetic space group corresponds to the magnetic point group $4'/mmm'$. 
In this ordered phase, time-reversal symmetry $\mathcal{T}$ is broken. However, the AFM CoF$_2$ crystal remains invariant under the combined operation of $90^\circ$-rotation $\rho$ (around the $z$-axis) followed by the time-reversal symmetry ($\rho \mathcal{T}$), giving rise to the altermagnetic nature of the ordered state. 
As a result of this specific symmetry, the order parameter is not the net magnetization (it is zero) nor a magnetic quadrupole (it is also zero), but a magnetic octupolar tensor of third-rank $\mathcal{O}^M_{ijk}$, defined as \cite{Bhowal}:
\begin{equation}
\mathcal{O}^M_{ijk} = \int r_i r_j m_k(\mathbf{r}) d^3r,
\label{eq:OM}
\end{equation}
where $\mathbf{m}(\mathbf{r})$ is the magnetization density and $\mathbf{r}$ is the position vector.
Figure~1b in the main text shows the angular distribution of the magnetization density for this ferroically ordered state \cite{Bhowal}.

Next we clarify the role of spin group symmetry, justify the use of magnetic point groups for the nonlinear optical tensor analysis, and explicitly connect the wave-vector dependence of the SHG process ($^c{\chi}^{(3)}:\mathbf{E}^{\omega} \nabla \mathbf{E}^{\omega}$) to the $\mathbf{k}$-dependent spin splitting that defines the altermagnetic state.

A proper description of CoF$_2$ requires distinguishing between two symmetry frameworks:

    \emph{1. Relativistic magnetic point group}. The magnetic point group $4'/mmm'$ considers all symmetries, including the spin-orbit coupling, and treats spins and spatial symmetries as locked. It accurately describes the macroscopic symmetry of magnetically ordered state, the ferroic ordering of magnetic octupoles. Importantly, this symmetry framework gives nonzero components of macroscopic tensors like the SHG susceptibility $^c\chi^{(3)}$ below $T_N$ presented in the main text, Table~I. This is the correct level for analyzing the existence and transformations of tensors in magnetically ordered crystals \cite{Birss,review,Pisarev,Fiebig1}.
    
    \emph{2. Non-relativistic spin group.} For CoF$_2$, the relevant spin point group is $^{\overline{1}}4/^{1}m^{1}m^{\overline{1}}m$ (group No. 195 in Litvin's classification), which is isomorphic to the magnetic point group $4'/mmm'$ but provides a richer representation theory for electronic states \cite{Schiff}. 
    In this symmetry framework including the limit of a weak spin-orbit coupling, the symmetry can be higher, spin rotations and spatial rotations can be considered independently. This is the natural framework for describing altermagnetism, as it allows for spin-split bands even without spin-orbit coupling \cite{Smejkal2022_0, Schiff}.

The ferrotype magnetic octupole $\mathcal{O}^M$ is the fundamental macroscopic order parameter that couples to the $c$-type nonlinear susceptibility $^c\chi^{(3)}(\mathcal{O}^M)$ in our SHG experiment. The tensor relation linking $\mathbf{\mathcal{O}}^M$ to $^c{\chi}^{(3)}$ is detailed in SI 4.
The ferrotype $\mathcal{O}^M$ ordering is a direct consequence for the electronic band structure because it induces a momentum-dependent spin splitting even in the absence of spin-orbit coupling. 
The symmetry of this splitting is manifested in CoF$_2$ as an antisymmetric energy splitting $\Delta E(\mathbf{k}) = E_{\uparrow}(\mathbf{k}) - E_{\downarrow}(\mathbf{k}) \propto k_xk_y $~\cite{Smejkal2022_0}, defining the signature of a $d$-wave altermagnet.

The altermagnetic spin splitting $\Delta E(\mathbf{k})$ depends on the electronic wave vector $\mathbf{k}$. Both phenomena are thus fundamentally linked 
through the law of conservation of angular momentum, determining the phenomenon of spatial dispersion \cite{Agranovich}.
The electric-quadrupole mechanism is naturally sensitive to the same anisotropic, $\mathbf{k}$-dependent electronic structure that defines the altermagnetic state, making SHG a good probe \cite{Ma}.
To quantitatively demonstrate the d-wave splitting in CoF$_2$, we employ a model based on the spin group analysis of Schiff et al. \cite{Schiff}. We consider a Hamiltonian for non-interacting fermions with Hund's exchange coupling to fixed classical localized moments in the AFM CoF$_2$ structure \cite{Schiff}:
\begin{equation}
H = -t_1\sum_{\langle i,j \rangle} c_{i\sigma}^{\dagger} c_{j\sigma} - \sum_{a=1,2} t_3^a \sum_{\langle\langle\langle i,j \rangle\rangle\rangle_a} c_{i\sigma}^{\dagger} c_{j\sigma} - J\sum_i c_{i\alpha}^{\dagger} \mathbf{h}_i \cdot \boldsymbol{\sigma}_{\alpha\beta} c_{i\beta},
\label{eq:hamiltonian}
\end{equation}
where $t_1$ is the nearest-neighbor hopping parameter, $t_3^1$, $t_3^2$ are the third-neighbor hopping parameters of two types, $J$ is the exchange integral, $\boldsymbol{\sigma}$ are the Pauli matrices, and
$\mathbf{h}_i = (0,0,\pm1)$ represent the localized magnetic moments on the two sublattices.
In momentum space, this Hamiltonian becomes block-diagonal in spin space, allowing for a pure spin-up and spin-down description (see Eqs.~(12) and (13) in \cite{Schiff}).

Our calculations (Figs.~1c,d in the main text) are in good agreement with large-scale ab-initio studies \cite{Guo, CoF2_ref}, see Figs.~2c. On basis of these calculations, CoF$_2$ can be considered as the $d$-wave altermagnet with a pronounced spin splitting of the electronic bands. 
The key features are: the presence of $d$-wave spin splitting without spin-orbit coupling, the antisymmetric energy splitting $\Delta E(\mathbf{k}) = -\Delta E(-\mathbf{k})$ in respect to the wave vector $\mathbf{k}$, the role of continuous rotational symmetry and the connection between spin group symmetry and band degeneracies. This large, $\mathbf{k}$-dependent spin splitting, related on the exchange interaction and not on the spin-orbit coupling, is the primary driver of altermagnetism in CoF$_2$. However,  the $\mathbf{k}$-dependent splitting is  not necessary large for other crystals with the same spin structure.
For example, AFM MnF$_2$ shares the same crystal and magnetic symmetry as CoF$_2$, but recent reports suggest its $d$-wave spin splitting is rather small \cite{Morano,Faure}. This highlights a critical point: fulfilling the symmetry requirements for altermagnetism is necessary but not sufficient for a large splitting. The magnitude depends on specific material parameters, such as the hopping parameters and exchange integrals. In CoF$_2$, these parameters conspire to produce a large $\mathbf{k}$-dependent spin splitting, making it a proper model system for optical studies of altermagnetism. The splitting on the order of hundreds of meV provides the microscopic foundation for the ferroic-octupolar SHG signal observed in CoF$_2$ below $T_N$. 
One would therefore expect the SHG response in MnF$_2$ having the same symmetry as CoF$_2$ in the spin-ordered phase, but with considerably weaker intensity, consistent with its much smaller band splitting. Important to add that the ground state of Mn$^{2+}$ is an orbital singlet whereas the ground state of Co$^{2+}$ is an orbital triplet which is strongly split by the spin-orbit coupling.

Since the macroscopic nonlinear tensor $^c\chi^{(3)}(\mathcal{O}^M)$ is defined by the magnetic point group $4'/mmm'$, the symmetry, magnitude and microscopic origin of the SHG effect, in particular its relation to the wave vector $\mathbf{k}$, have to be understood using the relativistic magnetic point group framework.

\section{Symmetry analysis of second-harmonic generation in CoF$_2$}

This section is devoted to a detailed symmetry analysis of second-harmonic generation (SHG) in centrosymmetric CoF$_2$. Based on the microscopic description of the altermagnetic state from SI~2, we now employ the language of nonlinear optics to justify the use of the magnetic point group $4'/mmm'$ for analyzing the nonlinear optical response and demonstrate how the ferroic-octupolar order parameter $\mathcal{O}^M$ enables SHG via an electric-quadrupolar mechanism. Finally, we explicitly link this mechanism to the wave-vector-dependent properties of the material.

In the general case, the nonlinear polarization at the second-harmonic frequency, $\mathbf{P}^{2\omega}$, induced by an electric field $\mathbf{E}^\omega$, is described by an expansion in multipoles. In the electric-dipole approximation (ED), $\mathbf{P}^{2\omega}$ can be written as:
\begin{equation}
\mathbf{P}^{2\omega}_{\text{ED}} = \epsilon_0\text{ }^i\chi^{(2)}_{\text{ED}}:\mathbf{E}^\omega \mathbf{E}^\omega,
\label{eq:ED}
\end{equation}
where $\epsilon_0$ is the vacuum permittivity.
This process, governed by the polar third-rank tensor $^i\chi^{(2)}_{\text{ED}}$, is strictly forbidden in crystals with a center of inversion, such as paramagnetic CoF$_2$ (point group $4/mmm$) \cite{Bloembergen,Shen,Boyd}.
Consequently, the SHG signal observed in our experiments must originate from higher-order multipolar mechanisms, which are symmetry-allowed even in centrosymmetric media. The two most relevant contributions are the magnetic-dipole (MD) and electric-quadrupole (EQ) terms \cite{Adler, Hoshi}:
\begin{equation}
\mathbf{P}^{2\omega}_{\text{MD}} = i \epsilon_0\text{ }{^i\chi}^{(2)}_{\text{MD}} : \mathbf{E}^\omega \mathbf{H}^\omega, 
 \label{eq:MD}
\end{equation}
\begin{equation}
\mathbf{P}^{2\omega}_{\text{EQ}} = \epsilon_0\text{ }{^i\chi}^{(3)}_{\text{EQ}} : \mathbf{E}^\omega \nabla \mathbf{E}^\omega, 
\label{eq:EQ}
\end{equation}
where $\mathbf{H}^\omega$ is the magnetic field of the light wave, and $\nabla \mathbf{E}^\omega$ represents the vector differential operator of the optical electric field, which introduces spatial dispersion, ${^i\chi}^{(2)}_{\text{MD}}$ is the axial third-rank $i$-type tensor, ${^i\chi}^{(3)}_{\text{EQ}}$ is the polar forth-rank $i$-type tensor.
However, as noted in the main text, for CoF$_2$ in its paramagnetic and AFM phases the specific non-zero components (e.g., $^i\chi_{yxz} = -^i\chi_{xyz}$ \cite{Birss}) allowed by the magnetic point group $4'/mmm'$ do not produce the SHG. 
Therefore, the dominant mechanism for the observed magnetic SHG signal is the EQ mechanism, described by Eq.~\ref{eq:EQ}.

A correct symmetry analysis of the SHG tensors for magnetically ordered crystals have to be considered according to the Neumann’s principle \cite{Neumann,Birss} by the magnetic point groups
rather than the magnetic space groups, also known as Shubnikov groups \cite{Shubnikov,Bardley}.
For systems where spin-orbit coupling is present, as it always is in real materials at the macroscopic level, the symmetry operations in real space and spin space are locked. 
It is important to clarify the distinction raised in the context of altermagnetism. 
While the microscopic origin of the spin splitting is best described by non-relativistic spin groups (as discussed in SI~2), the macroscopic physical property tensors that govern responses like SHG must transform according to the magnetic point group. 
The spin group is a higher symmetry that applies in the idealized cases with the weak or zero spin-orbit coupling \cite{Schiff}.
However, in a real crystal, the tensor describing light-matter interaction is constrained by the full relativistic symmetry of the magnetically ordered medium, hence the magnetic point group correctly captures the resulting macroscopic symmetry.
Therefore, our use of the tensors derived from $4'/mmm'$ (Table I, the main text) is both justified and standard practice in nonlinear optics of magnetic materials \cite{review,Pisarev,Fiebig1,Xiao}. 

Below $T_N$, the SHG signal in the electric-quadrupolar approximation is a superposition of a temperature-independent crystallographic contribution and a temperature-dependent spin-induced contribution. The total nonlinear polarization can be represented as the sum of two parts:
\begin{equation}
\mathbf{P}^{2\omega} = \epsilon_0 \text{ }^{i}\boldsymbol{\chi}^{(3)} :\mathbf{E}^\omega \nabla \mathbf{E}^\omega + i\epsilon_0 \text{ }^{c}\boldsymbol{\chi}^{(3)}(\mathcal{O}^M) :\mathbf{E}^\omega \nabla \mathbf{E}^\omega,
\label{eq:EQ_total}
\end{equation}
where the first term, governed by the $i$-type polar forth-rank tensor ${}^{i}{\chi}^{(3)}$, is invariant under time-reversal $\mathcal{T}$. It exists in both the paramagnetic and AFM phases and represents the crystallographic background SHG. The second term, governed by the $c$-type polar forth-rank tensor ${}^{c}{\chi}^{(3)}$, changes sign under $\mathcal{T}$ and is directly proportional to the ferroic-octupolar order parameter $\mathcal{O}^M$ \cite{Bhowal}. It emerges only below $T_N$ due to $\mathcal{T}$-symmetry breaking.

The non-zero tensor components for both ${}^{i}{\chi}^{(3)}$ (point groups $4/mmm$ and $4'/mmm'$) and ${}^{c}{\chi}^{(3)}$ (point group $4'/mmm'$) are listed in Table I of the main text. A key feature of the magnetic $c$-type tensor is the sign reversal between components such as $^c\chi_{xxxx} = -^c\chi_{yyyy}$, which directly reflects the $\rho\mathcal{T}$ symmetry of the spin-ordered state of CoF$_2$ and is responsible for the characteristic change in the SHG polarization patterns shown in Fig.~2 of the main text.

The electric-quadrupole mechanism explicitly involves the vector differential operator of the electric field, $\nabla \mathbf{E}^\omega$. For a monochromatic plane wave, this operator introduces a linear dependence on the light's wave vector $\mathbf{k}^0$, as $\nabla \mathbf{E}^\omega \propto i \mathbf{E}^\omega \mathbf{k}^0$. Equation~\ref{eq:EQ_total} can thus be rewritten to highlight this dependence:
\begin{equation}
\mathbf{P}^{2\omega} = i\epsilon_0\text{ }^i\chi^{(3)}:\mathbf{E}^\omega\mathbf{E}^\omega \mathbf{k}^0 + \epsilon_0 \text{ }^c\chi^{(3)}(\mathcal{O}^M):\mathbf{E}^\omega\mathbf{E}^\omega \mathbf{k}^0.
\label{eq:EQ_k}
\end{equation}
This formulation explicitly shows that the SHG intensity depends on the direction and magnitude of $\mathbf{k}^0$, a phenomenon known as spatial dispersion \cite{Agranovich}. The nonlinear components for Eq.~\ref{eq:EQ_k} are similar to those listed in Table~1 in the main text, a difference is that the nonlinear tensors in Eq.~\ref{eq:EQ_k} are symmetric with respect to the permutation of their second and third indices.

The wave-vector dependence of spin-splittings is a key feature of altermagnets. As shown in SI~2, the defining property of a $d$-wave altermagnet like CoF$_2$ is the strong, anisotropic splitting of electronic bands that depends on the electronic wave vector $\mathbf{k}$. Both the SHG process (through $\mathbf{k}^0$) and the altermagnetic band structure (through $\mathbf{k}$) are governed by wave-vector-dependent interactions.
During optical excitation, momentum conservation dictates that the wave vector $\mathbf{k}^0$ of the light field is transferred to the excited electron, directly coupling the macroscopic photon momentum to the microscopic electronic states. The electric-quadrupole mechanism, therefore, provides a natural and symmetry-allowed pathway for this momentum transfer, enabling the light field to couple to the $\mathbf{k}$-dependent spin texture of the altermagnet. Consequently, SHG serves as a direct probe of the momentum-space characteristics of the hidden order, as has been recently proposed for other altermagnets like RuO$_2$ using a non-collinear SHG technique \cite{Ma}.

A direct experimental consequence of the $\mathbf{k}^0$-dependence in Eq.~\ref{eq:EQ_k} is that the SHG signal should be strongly suppressed when the fundamental wave vector has no in-plane component. In our $xz$-cut sample, at normal incidence ($\varphi = 0^\circ$), $\mathbf{k}^0$ is parallel to the sample normal (the $y$-axis, following our coordinate convention) and does not effectively activate the in-plane tensor components listed in Table I. Conversely, tilting the sample to an oblique incidence angle (tilting around $z$-axis for the angle $\varphi = 30^\circ$, see Inset in Fig.~2 of the main text) introduces an in-plane component of $\mathbf{k}^0$, enabling these components and generating a strong SHG signal.

\begin{figure}[H]
    \centering
    \includegraphics[width=0.6\textwidth]{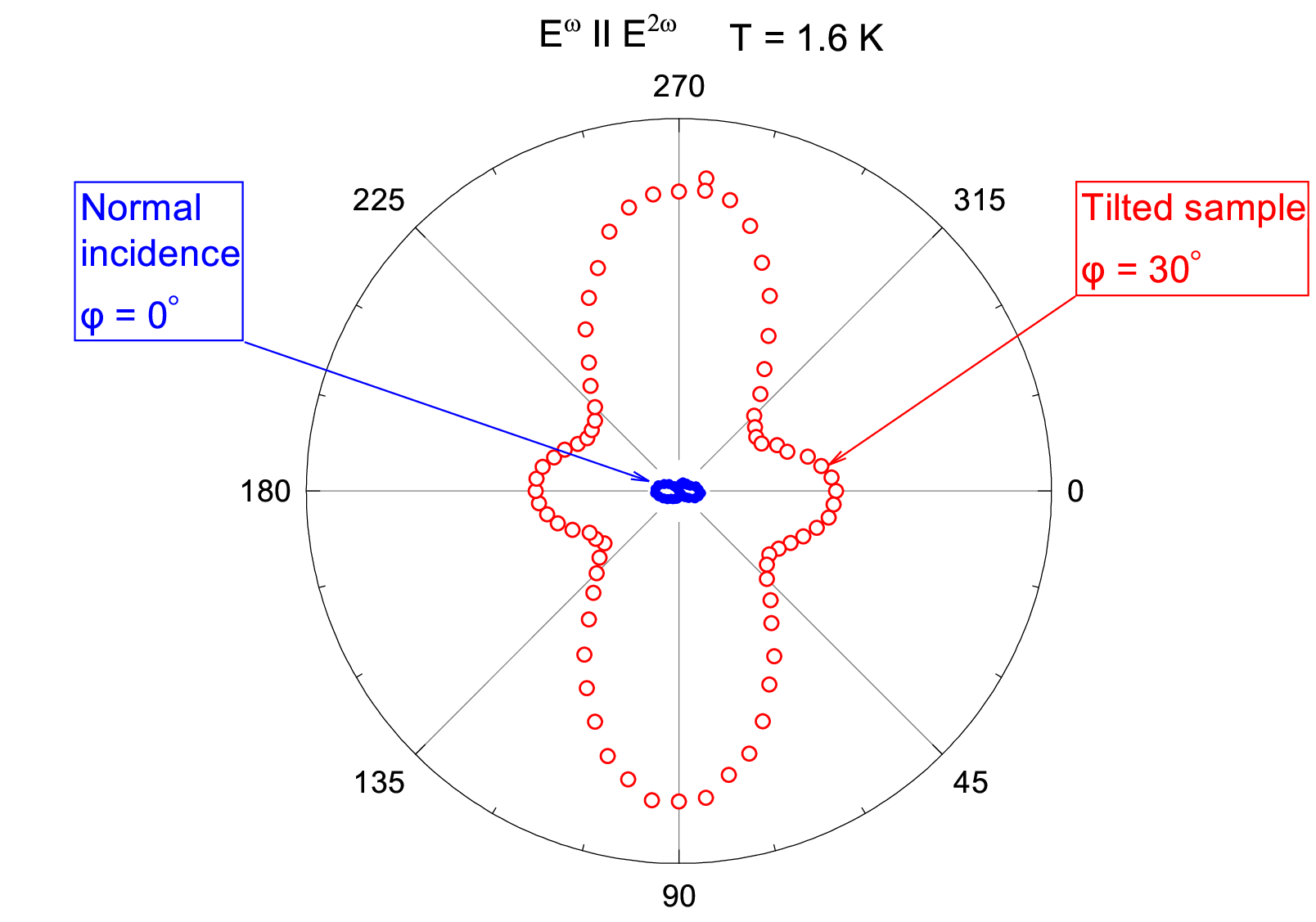}
    \caption{Polarization-dependent SHG recorded in the geometry $E^{\omega} \parallel E^{2\omega}$ at temperature of
1.6 K for the normal ($\varphi = 0^\circ$) and oblique incidence ($\varphi = 30^\circ$).}
    \label{fig:Norm_inc}
\end{figure}

We performed this exact control experiment, and the results are shown in Fig.~\ref{fig:Norm_inc}. At normal incidence, the SHG signal is vanishingly small (note the intensity scale is the same as in Fig.~2 of the main text). The finite, though tiny, residual signal can be attributed to higher-order effects, such as radial spatial dispersion arising from the tightly focused Gaussian beam used in the experiment \cite{Zeldovich, Kleinman, Chmela}. The dramatic increase in SHG intensity upon going to oblique incidence provides unambiguous proof that the observed signal is indeed driven by the $\nabla \mathbf{E}^\omega$ (or $\mathbf{k}^0$-dependent) electric-quadrupole mechanism, validating our symmetry analysis and its connection to the octupolar order parameter. 

To summarize the above, the three-photon nature of the SHG process allows for momentum resolution within the Brillouin zone through a non-zero momentum transfer. 
In principle, a non-collinear geometry, proposed in \cite{Ma} could selectively probe specific regions of the Brillouin zone, accessing up to $~0.6$~\% of its linear extent. However, in our collinear geometry, although the momentum transfer is approximately four times smaller, it remains sufficient to provide meaningful momentum sensitivity. 
The observed enhancement under oblique incidence confirms that the SHG signal couples to the wave-vector-dependent electronic structure. The symmetry analysis presented here, grounded in the magnetic point group $4'/mmm'$, fully accounts for the observed SHG polarization anisotropies and their temperature evolution. 
The explicit dependence on the light wave vector $\mathbf{k}^0$ not only confirms the electric-quadrupole nature of the effect but also establishes SHG as a momentum-sensitive probe ideally suited to optically investigate the $\mathbf{k}$-dependent electronic structure that defines the altermagnetic phase.

\section{Magnetic octupole moments and their relation to SHG}
\begin{table}[h]
\centering
\caption{\label{tab:table01} Nonzero components of the nonlinear susceptibility $^i\chi_{ijklmno}$ relating octupolar components $^c\mathcal{O}_{mno}$ and the nonlinear susceptibility $^c\chi_{ijkl}$}

\begin{tabular}{cccccc}\hhline{======}
\hline
$^i\chi_{ijklmno}$ ($^c\mathcal{O}_{zyx}  =\text{ }^c\mathcal{O}_{zxy}$) \\
 $xxyyzyx = -yyxxzxy$, $xxxxzyx = -yyyyzxy$, $yyyyzyx = -xxxxzxy$, $zyzyzxy = -zxzxzyx$ \\ \hline
 $^i\chi_{ijklmno}$ ($^c\mathcal{O}_{xzy} =\text{ }^c\mathcal{O}_{yzx} =\text{ }^c\mathcal{O}_{xyz} =\text{ }^c\mathcal{O}_{yxz}$) \\
 $xzxzxzy = -yzyzyzx$, $xxyyxzy = -yyxxyzx$, $xzxzyzx = -yzyzxzy$, \\ 
 $xxxxyxz = -yyyyxyz$, $xzxzxyz = -yzyzyxz$, $yxyxyxz = -xyxyxyz$, \\
 $xxyyxyz = -yyxxyxz$, $yxxyyxz = -xyyxxyz$, $yyyyyxz = -xxxxxyz$ \\ \hline
\end{tabular}
\end{table}

The magnetic octupole moment $\mathcal{O}^M$ is directly related to the nonlinear polarization responsible for the observed SHG in CoF$_2$:
\begin{equation}
\mathbf{P}^{2\omega} = \mathrm{i}\epsilon_0\text{ }^c\chi^{(3)}(\mathcal{O}^M):\mathbf{E}^{\omega}\nabla \mathbf{E}^{\omega}.
\end{equation}
The magnitude of $\mathcal{O}^M$ provides a quantitative measure of the altermagnetic order parameter that drives the SHG response below $T_N$.

The relationship between the nonlinear susceptibility $^c\chi_{ijkl}$ and the magnetic octupole moment $^c\mathcal{O}_{mno}$ in CoF$_2$ is given by:
\begin{equation}
^c\chi_{ijkl} =\text{ }^i\chi_{ijklmno} :\text{ }^c\mathcal{O}_{mno}
\label{eq:MO_khi}
\end{equation}
where $^i\chi_{ijklmno}$ is an axial $i$-type tensor transforming under the crystallographic point group $4/mmm$ and the magnetic point group $4'/mmm'$, and $^c\mathcal{O}_{mno}$ is an axial $c$-type tensor transforming under the magnetic point group $4'/mmm'$. 
The resulting fourth-rank tensor $^c\chi_{ijkl}$ therefore inherits the transformation properties of the magnetic octupole and is of $c$-type.
For the magnetic point group $4'/mmm'$, nonzero components of the magnetic octupole tensor $^c\mathcal{O}_{mno}$ are
 $zyx  = zxy$ and  $xzy = yzx = xyz = yxz$.

Table~1 presents the nonzero components of the sevens-rank tensor $^i\chi_{ijklmno}$. These components contribute to all independent components of $^c\chi_{ijkl}$ reported in the main text.
Thus, the components presented in Table~1 constitute the complete set that governs the ferrotype magnetic octupolar contribution to $^c\chi_{ijkl}$ in CoF$_2$. Consequently, they enable a full description of the second-harmonic generation induced by the altermagnetic order parameter $\mathcal{O}^M$.

\section{Modeling of temperature and polarization dependencies of SHG}

Based on the symmetry analysis (Eq. \ref{eq:EQ_total}) and the relation linking the nonlinear susceptibility to the magnetic octupole (Eq. \ref{eq:MO_khi}), we can now model the experimental SHG data, which are the polarization anisotropies and temperature dependencies of SHG intensity. In this section, we construct explicit phenomenological expressions for the nonlinear polarization $P^{2\omega}$ based on the nonzero components of crystallographic $^i{\chi}^{(3)}$ and ferroic-octupole-induced $^c{\chi}^{(3)}(\mathcal{O}^M)$. By fitting these expressions to the experimental data for different polarization geometries (parallel $E^{\omega} \parallel E^{2\omega}$ and crossed $E^{\omega} \perp E^{2\omega}$) as a function of the azimuthal angle $\theta$, we can obtain the relative magnitudes of the tensor coefficients. Furthermore, by comparing the temperature evolution of the extracted magnetic contribution with the independently known behavior of the squared spin order parameter of CoF$_2$ obtained from neutron spectroscopy data \cite{Chatterji}, we can confirm its spin-related ordering. This modeling therefore provides the necessary link between the tensor formalism and SHG experimental data.

The nonlinear polarizations that generate SHG in our experimental geometry are obtained by projecting the general tensor expansions onto the polarization directions of the fundamental and second‑harmonic waves. To model the SHG intensity as a function of the azimuthal angle $\theta$ and the angle of sample tilting $\varphi$, we construct explicit expressions for the nonlinear polarization $P^{2\omega}$ based on the nonzero tensor components listed in Table I of the main text. For the parallel geometry ($E^{\omega} \parallel E^{2\omega}$), the crystallographic and magnetic contributions to the nonlinear polarization are given by:
\begin{equation}
\begin{aligned}
P^{crys}_{\|} = &-\sin\varphi \sin\theta \left( a_1 \sin\varphi \cos\varphi \sin^2\theta + b_1 \cos^2\varphi \sin^2\theta + c_1 \sin^2\varphi \sin^2\theta - d_1 \sin\varphi \sin\theta \cos\theta + e_1 \cos^2\theta \right) \\
&-\cos\varphi \sin\theta \left( a_1 \sin\varphi \cos\varphi \sin^2\theta + b_1 \sin^2\varphi \sin^2\theta + c_1 \cos^2\varphi \sin^2\theta - d_1 \cos\varphi \sin\theta \cos\theta + e_1 \cos^2\theta \right) \\
&+\cos\theta \left( -f_1 \cos\varphi \sin\theta \cos\theta - f_1 \sin\varphi \sin\theta \cos\theta + g_1 \cos^2\varphi \sin^2\theta + g_1 \sin^2\varphi \sin^2\theta + h_1 \cos^2\theta \right),
\end{aligned}
\label{eq:PC_parr}
\end{equation}
\begin{equation}
\begin{aligned}
P^{magn}_{\|} = &-\sin\varphi \sin\theta \left( a_2 \sin\varphi \cos\varphi \sin^2\theta + b_2 \cos^2\varphi \sin^2\theta + c_2 \sin^2\varphi \sin^2\theta - d_2 \sin\varphi \sin\theta \cos\theta + e_2 \cos^2\theta \right) \\
&+\cos\varphi \sin\theta \left( a_2 \sin\varphi \cos\varphi \sin^2\theta + b_2 \sin^2\varphi \sin^2\theta + c_2 \cos^2\varphi \sin^2\theta - d_2 \cos\varphi \sin\theta \cos\theta + e_2 \cos^2\theta \right) \\
&+\cos\theta \left( f_2 (\sin\varphi - \cos\varphi) \sin\theta \cos\theta + g_2 (\cos^2\varphi - \sin^2\varphi) \sin^2\theta \right),
\end{aligned}
\label{eq:PM_parr}
\end{equation}
where the parameters $a_1,b_1,c_1,d_1,e_1,f_1,g_1,h_1$ correspond to the independent nonzero components of the crystallographic $i$-type tensor ${}^{i}\chi^{(3)}$, and $a_2, b_2, c_2, d_2, e_2, f_2, g_2$ correspond to the components of the magnetic $c$-type tensor ${}^{c}\chi^{(3)}$, listed respectively in Table I of the main text.

For the crossed geometry ($E^{\omega} \perp E^{2\omega}$), the corresponding expressions are:
\begin{equation}
\begin{aligned}
P^{crys}_{\perp} = &-\sin\varphi \cos\theta \left( a_1 \sin\varphi \cos\varphi \sin^2\theta + b_1 \cos^2\varphi \sin^2\theta + c_1 \sin^2\varphi \sin^2\theta - d_1 \sin\varphi \sin\theta \cos\theta + e_1 \cos^2\theta \right) \\
&-\cos\varphi \cos\theta \left( a_1 \sin\varphi \cos\varphi \sin^2\theta + b_1 \sin^2\varphi \sin^2\theta + c_1 \cos^2\varphi \sin^2\theta - d_1 \cos\varphi \sin\theta \cos\theta + e_1 \cos^2\theta \right) \\
&-\sin\theta \left( -f_1 \cos\varphi \sin\theta \cos\theta - f_1 \sin\varphi \sin\theta \cos\theta + g_1 \cos^2\varphi \sin^2\theta + g_1 \sin^2\varphi \sin^2\theta + h_1 \cos^2\theta \right),
\end{aligned}
\label{eq:PC_cross}
\end{equation}
\begin{equation}
\begin{aligned}
P^{magn}_{\perp} = &-\sin\varphi \cos\theta \left( a_2 \sin\varphi \cos\varphi \sin^2\theta + b_2 \cos^2\varphi \sin^2\theta + c_2 \sin^2\varphi \sin^2\theta - d_2 \sin\varphi \sin\theta \cos\theta + e_2 \cos^2\theta \right) \\
&+\cos\varphi \cos\theta \left( a_2 \sin\varphi \cos\varphi \sin^2\theta + b_2 \sin^2\varphi \sin^2\theta + c_2 \cos^2\varphi \sin^2\theta - d_2 \cos\varphi \sin\theta \cos\theta + e_2 \cos^2\theta \right) \\
&-\sin\theta \left( f_2 (\sin\varphi - \cos\varphi) \sin\theta \cos\theta + g_2 (\cos^2\varphi - \sin^2\varphi) \sin^2\theta \right),
\end{aligned}
\label{eq:PM_cross}
\end{equation}
For both geometries, the azimuthal angle $\theta$ is measured from the $z$-direction of the CoF$_2$ crystal for the $E^{\omega}$ polarization (see inset in Fig.~2 of the main text).

Above the Néel temperature $T_N = 38$~K the magnetic order parameter $\mathcal{O}^M$ vanishes, and consequently the $c$-type tensor $^c\chi^{(3)}(\mathcal{O}^M)\equiv0$. The SHG signal in the paramagnetic phase is therefore purely crystallographic, described by the time-even tensor $^i\chi^{(3)}$. Below $T_N$, the magnetic contribution appears and its magnitude grows with the altermagnetic order. 
The total SHG intensity signal consists of the temperature-independent crystallographic and temperature-dependent spin-induced contributions:
\begin{equation}
  I^{2\omega} \propto |\mathbf{P}^{crys} + \mathbf{P}^{magn}|^2 \propto \big |\big(^i\chi^{(3)}+ i\text{ }^c\chi^{(3)}(\mathbf{\mathcal{O}}^M)\big)I^{\omega}\big |^2 = |^i\chi^{(3)}I^{\omega}|^2 + |^c\chi^{(3)}(\mathbf{\mathcal{O}}^M)I^{\omega}|^2.
\label{eq:I_2w}
\end{equation} 
An important consequence of Eq.~\ref{eq:I_2w} is the absence of interference between the crystallographic and magnetic contributions. The factor $i$ in front of the magnetic term reflects the $\pi/2$ phase shift between the time-even ($i$-type) and time-odd ($c$-type) nonlinear susceptibilities. This quadrature relationship means that the two contributions add in intensity  and  the cross term does nor appear. 
The absence of the interference of $i$-type and $c$-type nonlinear contributions to the SHG intensity has an advantage in our study. The SHG signal measures the sum of squares $|^i\chi^{(3)}|^2 + |^c\chi^{(3)}(\mathcal{O}^M)|^2$, it is insensitive to the sign of $\mathcal{O}^M$. Important to say that the SHG signal therefore averages over domains constructively, yielding a robust measure of the order parameter rather than its orientation. This simplifies the interpretation of the temperature dependence and confirms that the observed SHG below $T_N$ directly tracks the growth of the spin order parameter.

\begin{table}[h!]
\centering
\caption{Crystallographic ($crys$) and spin-induced ($magn$) nonlinear polarizations for different geometries, different azimuthal angles $\theta$ 
and the angle of sample tilting $\varphi =30^\circ$.}
\label{tab:selection_rules2}
\begin{tabular}{@{}llll}\hhline{====}
& $\theta=0^\circ$ & $\theta=45^\circ$ & $\theta=90^\circ$ \\
\midrule
$P^{crys}_{\|}$ & $h_1$ & $\big(((a_1 + b_1 + 3c_1 + 4e_1 + 4f_1)\sqrt{3}+3a_1 + 3b_1+$ & $\big((a_1 + b_1 + 3c_1)\sqrt{3}+$ \\
&  & $+ c_1 + 8d_1 + 4e_1 + 8g_1 + 8h_1 + 4f_1)\big)\sqrt{2}/32$ & $+ 3a_1+ 3b_1+ c_1\big)/8$ \\ \hline
$P^{crys}_{\perp}$ & $-e_1(1 + \sqrt{3})/2$ & $-\big(((a_1 + b_1 + 3c_1 + 4e_1 - 4f_1)\sqrt{3} + 3a_1 + 3b_1+$ & $g_1$ \\
&  &  $+ c_1 + 8d_1 + 4e_1 - 8g_1 - 8h_1 - 4f_1)\sqrt{2}\big)/32$ &  \\ \hline
$P^{magn}_{\|}$ & 0 & $-\big((a_2 - b_2 - 3c_2 - 4e_2 + 4f_2)\sqrt{3} - 3a_2 + 3b_2+$ & $\big((-a_2 + b_2 + 3c_2)\sqrt{3}+$\\
 & & $+ c_2 - 4d_2 + 4e_2 + 4g_2 - 4f_2\big)\sqrt{2}/32$ & $ + 3a_2 - 3b_2 - c_2\big)/8$\\ \hline
$P^{magn}_{\perp}$ & $e_2(1-\sqrt{3})/2$ & $((a_2 - b_2 - 3c_2 - 4e_2 - 4f_2)\sqrt{3} - 3a_2 + 3b_2+$ & $-g_2/2$\\
&  &  $+ c_2 - 4d_2 + 4e_2 - 4g_2 + 4f_2)\sqrt{2}/32$ &  \\
\bottomrule &  &  &
\end{tabular}
\end{table}

Using the expressions (Eqs.~\ref{eq:PC_parr} -- \ref{eq:I_2w}), we performed a simultaneous fit to the experimental polarization dependencies measured at different temperatures for both parallel and crossed geometries. Table~\ref{tab:selection_rules2} lists the simplified expressions for the nonlinear polarization at selected azimuthal angles $\theta = 0^\circ$, $45^\circ$, and $90^\circ$, for a fixed sample tilting by $\varphi = 30^\circ$. These expressions were obtained by substituting the corresponding angles into Eqs.~(\ref{eq:PC_parr}) -- (\ref{eq:I_2w}). As seen from the Table~2, measurements at different $\theta$ values selectively emphasize different combinations of tensor components, allowing their relative magnitudes to be determined unambiguously from the global fit.

\begin{figure}[H]
    \centering
    \includegraphics[width=0.59\textwidth]{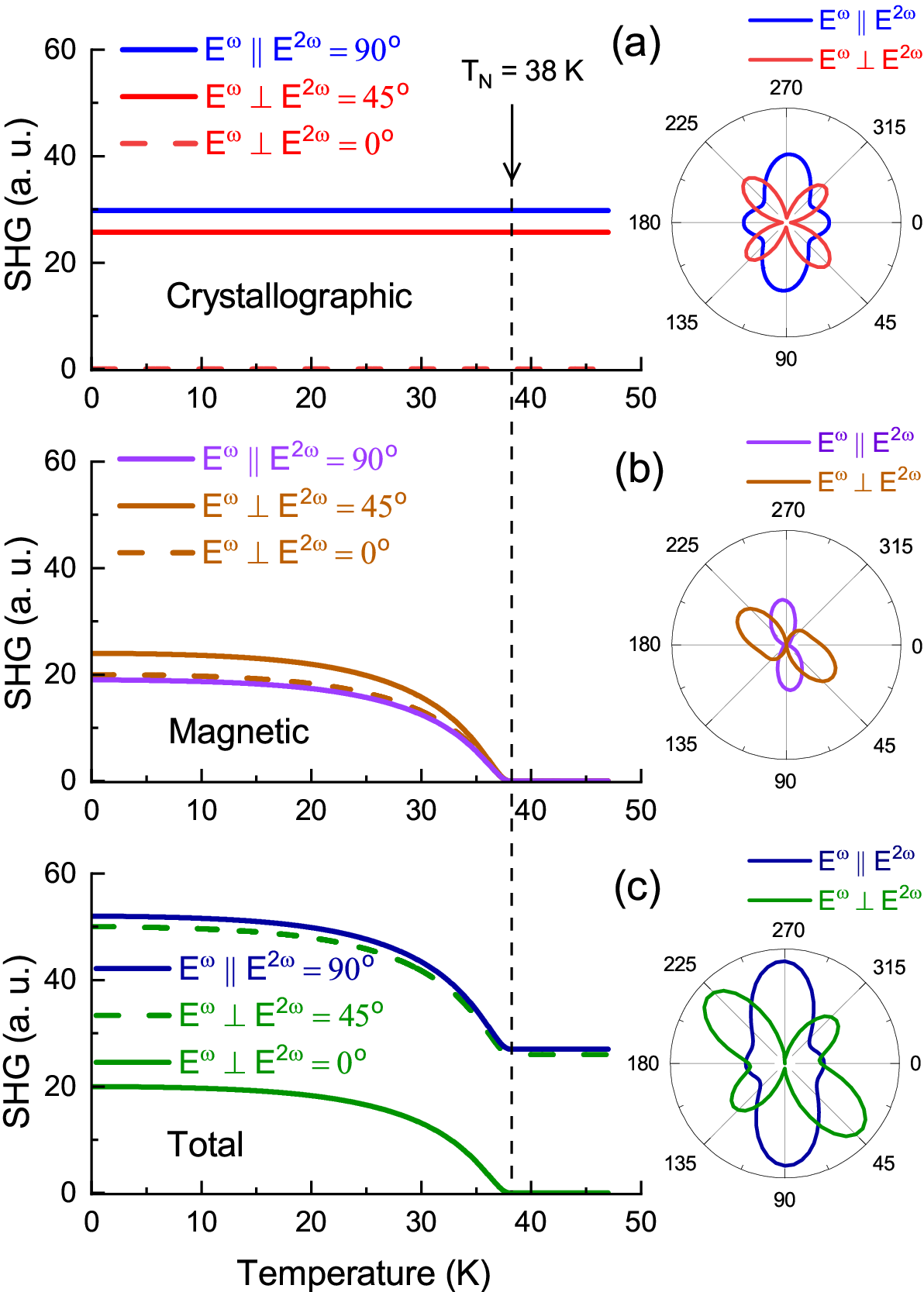}
    \caption{Temperature and polarization dependencies of SHG intensities modeled for: (a) the crystallographic contribution, (b) the spin-induced contribution (b) and (c) the total signal.}
    \label{SHG_model}
\end{figure}

The magnetic contribution was scaled by the squared ferroic-octupolar order parameter $\mathcal{O}^M(T)$, using its temperature dependence as independently determined from neutron spectroscopy data \cite{Chatterji}. The resulting model shows excellent agreement with the experimentally observed temperature dependencies presented in Fig.~3 of the main text, following the temperature behaviour of the squared order parameter.
Figure \ref{SHG_model} presents the results of this modeling. 
Panel (a) shows the crystallographic contribution, which is temperature-independent both above and below $T_N$. Its polarization patterns exhibit a fourfold symmetry in the crossed geometry and a twofold symmetry in the parallel geometry, consistent with the symmetry of the $i$-type tensor $^i\chi^{(3)}$ for point group $4/mmm$. 
Panel (b) displays the magnetic contribution, which vanishes above $T_N$ and grows below it following the squared order parameter. Its polarization patterns show a characteristic twofold symmetry in both geometries, reflecting the $c$-type tensor $^c\chi^{(3)}(\mathcal{O}^M)$ for the magnetic point group $4'/mmm'$. 
Panel (c) shows the total SHG signal as the sum of the two contributions (Eq.~\ref{eq:I_2w}), which reproduces all key features of the experimental data presented in Figs.~2 and 3 of the main text: the temperature-independent offset above $T_N$, the monotonic increase below $T_N$, and the evolution of the polarization anisotropies with temperature.

The excellent agreement between the modeled curves and the experimental data confirms the validity of our phenomenological approach. The results confirm that the magnetic contribution scales with the altermagnetic order parameter $\mathcal{O}^M$ and follows the symmetry dictated by the magnetic point group $4'/mmm'$. This provides direct experimental support for the electric-quadrupole mechanism and the symmetry analysis presented in the main text.

\section{Three-photon resonance conditions for optimizing the SHG response in CoF$_2$}

The symmetry analysis in Section 3 confirms the higher-order multipole nature of the process, consequently, the SHG signal is inherently weak, as expected for a centrosymmetric material. Optimizing the experimental conditions therefore requires careful selection of the fundamental photon energy to maximize the resonant enhancement of $^{i,c}\chi^{(3)}$ while avoiding strong absorption losses.

The optical absorption in CoF$_2$ across the near-infrared to visible range is dominated by the weak intra-ion $d$–$d$ transitions within the Co$^{2+}$ ($3d^7$) ions \cite{Barreda}. The relevant energy levels are shown in the Tanabe-Sugano diagram in Fig.~4a of the main text, which depicts the $3d^7$ electron configuration in the octahedral crystal field of the F$^{-}$ ions. The fundamental bandgap of CoF$_2$ is $5.5$~eV \cite{Formisano}, placing all $d$–$d$ transitions well within the transparency range of the crystal.

To analyze the resonance enhancement of $^{i,c}\chi^{(3)}$ quantitatively, we consider microscopic expressions for the nonlinear susceptibilities. For the two Co$^{2+}$ ions in the unit cell, and keeping terms resonant at both one- and two-photon frequencies, the susceptibility can be expressed as \cite{Hanamura,Fiebig1}:
\begin{equation} \label{eq3}
^{i,c}\chi^{(3)} \propto \sum_{i} \rho_{i}\sum_{m,k}\frac{(PP\nabla P)_{imki}}{(\hbar\omega_{mi}-2\hbar\omega -i\Gamma_{mi})(\hbar\omega_{ki}-\hbar\omega-i\Gamma_{ki})},
\end{equation}
where $\rho_{i}$ accounts for the thermal distribution within the ground-state multiplet of Co$^{2+}$, ${(PP\nabla P)_{imki}}$ are the matrix elements contributing to the nonlinear susceptibilities. The energies $\hbar\omega_{ki}$ and $\hbar\omega_{mi}$ denote the differences between the ground state $| i \rangle$ ($^4T_1$) and the intermediate state $| k \rangle$ ($^4T_2$), and between $| i \rangle$ and the excited state $| m \rangle$ ($^4A_2$, $^{4}T^*_{1}$), respectively. Based on the energy level scheme, the relevant intermediate states $|k\rangle$ for one-photon resonance near $1.05$ eV are the $^4T_2$ states, while the relevant excited states $|m\rangle$ for two-photon resonance near $2.1$ eV include $^4A_2$, $^{4}T^*_{1}$, and higher-lying levels. The damping parameters $\Gamma_{ki}$ and $\Gamma_{mi}$ broaden these resonances. The numerator includes products of electric-dipole matrix elements and terms representing the quadrupolar interaction, reflecting the $\nabla \mathbf{E}^\omega$ coupling in  Eq.~\ref{eq:EQ_total}.

Resonant enhancement occurs when either denominator approaches zero. Importantly, the $i$-type (crystallographic) and $c$-type (magnetic) susceptibilities correspond to the real and imaginary parts of this expression, respectively \cite{Hanamura,Fiebig1}. This distinction reflects their different physical origins: $^i\chi^{(3)}$ arises from virtual transitions within the parity-even sector, while $^c\chi^{(3)}(\mathcal{O}^M)$ is enabled by broken time-reversal symmetry and scales with the magnetic octupole order parameter.

In practice, the choice of photon energy involves a trade-off. Tuning $\hbar\omega$ into resonance with an intermediate state $|k\rangle$ enhances the SHG response via the one-photon denominator in Eq.~\ref{eq3}, but also increases absorption of the fundamental beam, risking sample heating or damage. Conversely, tuning $2\hbar\omega$ into resonance with an excited state $|m\rangle$ enhances the signal via the two-photon denominator, but strong absorption at the SHG frequency attenuates the detected photons, reducing sensitivity. The optimal working point balances these competing effects.
\begin{table}[h!]
\centering
\caption{Selection rules for the three-photon processes in CoF$_2$. Notation: $\checkmark$ – allowed, $o$ – weakly allowed via spin-orbit mixing.}
\label{tab:selection_rules}
\begin{tabular}{@{}lllll}\hhline{=====}
 & Process & Spin selection & Possible mechanism & Efficiency \\
\midrule 1 & $^{4}T_{1}\xrightarrow{\omega} {}^{4}T_{2}\xrightarrow{\omega} {}^{4}A_{2}\xrightarrow{2\omega} {}^{4}T_{1}$ & $\checkmark $ ($\Delta S=0$) &
Electric quadrupole &  $\sim 1$ \\
2 & $^{4}T_{1}\xrightarrow{\omega} {}^{4}T_{2}\xrightarrow{\omega} {}^{4}T^*_{1}\xrightarrow{2\omega} {}^{4}T_{1}$ & $\checkmark $ ($\Delta S=0$) &
Electric quadrupole &  $\sim 1$ \\
3 & $^{4}T_{1}\xrightarrow{\omega} {}^{4}T_{2}\xrightarrow{\omega} {}^{2}A_{1} \xrightarrow{2\omega} {}^{4}T_{1}$ & $o$ ($\Delta S=1$ twice) &
Spin-orbit mixing & $\sim 10^{-3}-10^{-4}$ \\
4 & $^{4}T_{1}\xrightarrow{\omega} {}^{4}T_{2}\xrightarrow{\omega} {}^{2}A_{2} \xrightarrow{2\omega} {}^{4}T_{1}$ & $o$ ($\Delta S=1$ twice) &
Spin-orbit mixing & $\sim 10^{-3}-10^{-4}$ \\
\bottomrule &  &  &  &
\end{tabular}%
\end{table}

To identify which specific electronic states dominate the SHG response under our experimental conditions, we now examine the possible three-photon pathways consistent with the $d$-$d$ energy level scheme of Co$^{2+}$. Table~3 summarizes the selection rules and relative efficiencies for the most relevant processes, allowing us to pinpoint the dominant resonance channel.

The selection rules summarized in Table 3 reveal that the SHG in CoF$_2$ is dominated by the electric-quadrupole processes 1 and 2, which are $3$–$4$ orders of magnitude more efficient than the spin-forbidden alternatives. Both processes share the same two-photon resonance with the $^4A_2$ and $^{4}T^*_{1}$ excited states. This explains the strong resonant enhancement observed at $2\hbar\omega = 2.1$ eV and confirms that the experimental conditions are optimally tuned to the $^4T_1 \rightarrow {}^4T_2 \rightarrow {}^4A_2(^{4}T^*_{1}) \rightarrow {}^4T_1$ pathway, as discussed in the main text.

\noindent\hrulefill